\documentclass{aastex}     
\usepackage{spr-astr-addons}  
\usepackage{url}
\usepackage{comment}
\usepackage{epstopdf}
\usepackage{gensymb}
\usepackage{xcolor}
\usepackage[caption = false]{subfig}
\usepackage{csvsimple, longtable, booktabs}

\usepackage[maxfloats=40] {morefloats} 

\begin{document}
\title{Vilnius Photometry and Gaia Astrometry of Melotte 105}
\shorttitle{Vilnius Photometry {\color{black}and Gaia Astrometry of Melotte 105}}
\shortauthors{<Banks, T.>}


\author{Timothy Banks\altaffilmark{1}}
\email{tim.banks\@nielsen.com}
{\color{black} \author{Talar Yontan\altaffilmark{2}}}
 \author{Sel\c{c}uk Bilir\altaffilmark{3}}
{\color{black} \author{Remziye Canbay\altaffilmark{2}}}

\altaffiltext{1}{Nielsen, Data Science, 200 W Jackson Blvd \#17, Chicago, IL 60606, USA. Email: tim.banks@nielsen.com}
{\color{black} \altaffiltext{2}{Istanbul University, Institute of Graduate Studies in Science, Programme of Astronomy and Space Sciences, 34116, Beyaz{\i}t, Istanbul, Turkey}}
{\color{black} \altaffiltext{3}{Istanbul University, Faculty of Science, Department of Astronomy and Space Sciences, 34119, University, Istanbul, Turkey}}

\vspace{2mm}

\begin{abstract}

Archival Vilnius CCD photometric observations are presented for the heavily reddened star cluster Melotte 105, resulting in colour-magnitude diagrams and spectral class estimates. There is considerable lack of agreement between studies for reddening, age, and distance for this cluster explaining why the archival data are being made available by this paper. The derived reddening $E(B-V)=0.34\pm0.04$ mag and the distance $V-M_{V}=12.9\pm0.3$ mag 
{\color{black}  directly
from the Vilnius photometry. The {\em Gaia} Data Release 2 (DR2) and Vilnius 
photometric data of the cluster were used to estimate the structural parameters of the cluster, probability of stellar membership in the cluster, the distance modulus and the cluster age. Lack of $Y$ band observations prevented determination of metal abundance. 
The values of the colour excess and distance module are determined by two different methods (i.e., Q and Zero Age Main Sequence, or ZAMS, methods).
A distance modulus of $12.85\pm0.07$ mag was derived by ZAMS fitting, in good agreement with the above estimate. ZAMS fitting indicates a reddening of $0.403 \pm 0.02$ mag, within two sigma of the estimate above. The cluster's metallicity and age are estimated to be 0.24 dex and $240\pm25$ Myr, respectively. The derived mass function is in good agreement with the Salpeter slope. The cluster space velocity components $(U, V, W)$ were determined as ($-3.90\pm3.34$, $-13.76\pm5.69$, $+3.45\pm0.41$) km/s. Perigalactic and apogalactic distances were obtained as $R_{p}=6.85$ and $R_{a}=7.44$ kpc respectively. The maximum vertical distance from the Galactic plane was calculated as $Z_{max}=84$ pc and the eccentricity of the orbit was determined as $e=0.042$.}
      
\end{abstract}



\section{Introduction}

The seven filter intermediate band Vilnius system (see Strai\v zys
1992a, Forbes 1996) makes possible the purely photometric determination
of the spectral classes, absolute magnitudes, and metallicities of
stars while also correcting for interstellar reddening. This is
facilitated by the careful thought given to the positioning and widths
of the filters relative to the spectral features of all luminosity
classes. For example, the $U$ filter measures the ultraviolet intensity
below the Balmer jump, while the $P$ filter is placed on the jump itself,
allowing luminosity determinations for early type stars. The $X$ filter
measures a wavelength range longer than the Balmer jump, between the
$\rm H_{\delta}$ and $\rm H_{\epsilon}$ lines. This filter measures the
continuum intensity after the Balmer jump for early-type stars, and the
metallic-line blanketing for late-type stars. $Y$ is centred around 466
nm, excluding the interstellar band at 433 nm and the $\rm H_{\gamma}$
line. The $Z$, $V$ and $S$ filters coincide with features in late-type stars
(see Figure 1 of Dodd, Forbes \& Sullivan, 1993). $Z$ is near the bottom
of an absorption feature in late-type stars. The absorption
intensity of the Mg II triplet lines and the Mg H band is measured. $V$
measures the continuum at nearly the mean wavelength of the Johnson $V$
filter. In conjunction with the $Y$ filter, a direct conversion may be
made to the Johnson $BV$ system (Forbes 1996). $S$ is centred on the
$\rm H_{\alpha}$ line at 656 nm. It is used to separate the B emission
(Be) stars from normal B stars. The filter measures the absorption or
emission intensity of the $\rm H_{\alpha}$ line in early-type stars, or the
pseudo-continuum (due to metallic lines) in late-type stars. Unlike the
other filters, which were\footnote{At the time of data collection for
  this paper.} Soviet-made coloured glasses cemented with
Canadian Balsam, the $S$ filter is an interference filter.  Peculiar
stars, such as metal-deficient giants and blue horizontal branch
stars, can be recognised using the two and three dimensional
classification schemes of the system (Strai\v zys 1992a,b), making the
filter set well suited for the study of star clusters. Further details
of the rationale in the design of the filter set may be found in
Strai\v zys \& Sviderskien\. e (1972).

The Vilnius colour indices $U-P$, $P-X$, $X-Y$, $Y-Z$, $Z-Y$, and $V-S$ 
are set to zero for unreddened O-type stars. Colours for all normal stars are
therefore positive. In light of these capabilities, the idea of extending the 
system to the southern hemisphere was first considered in 1985 at the Royal
Observatory Edinburgh, given that none of the already established
standard regions (e.g. Zdanavi\v cus et al. 1969, \v Cernies et
al. 1989, and \v Cernies \& Jasevi\v cus 1992) extended south of the
celestial equator. This programme commenced in 1988 using the 61cm
telescopes at Mount John University Observatory (MJUO), with the
initial goal of establishing standards near the South Celestial Pole
and also bright ($V<7$ mag) stars generally distributed south of $-20$
degrees. Many southern Vilnius standard stars have been established in
NGC 4755. Further details on the programme may be found in Forbes
(1993) and Dodd, Forbes~\& Sullivan (1993).

\begin{table*}[tb]
\centering
\caption{Transformation co-efficients where $\rm m_1$, $\rm m_2$, 
and $\rm m_3$ are as given in equation 1. `Std RMS' gives the expected 
RMS given the formal uncertainties listed by Forbes (1996) for the 
standard stars, while `RMS' is the root mean square error from the 
fits themselves. The column `\#' indicates the number of standard 
star observations made per filter. 15 individual standards were 
observed, with a ($Y-V$) colour range of 0.31 to 1.71. The airmass 
range was 1.0 to 1.8.}
\begin{tabular}{|c|c|c|c|c|c|r|}
\hline
Filter & $\rm m_1$ & $\rm m_2$ & $\rm m_{3}$ & RMS & Std RMS & \# \\
\hline
$U$ & $17.531 \pm 0.030$ & $-0.008 \pm 0.007 $ & $0.654 \pm 0.019$ & 0.031 & 0.027 & 40 \\
$P$ & $17.260 \pm 0.031$ & $+0.091 \pm 0.006$ & $0.555 \pm 0.019$ & 0.032 & 0.023 & 47 \\
$X$ & $17.488 \pm 0.028$ & $+0.015 \pm 0.010$ & $0.365 \pm 0.017$ & 0.030 & 0.018 & 50 \\
$Z$ & $17.622 \pm 0.025 $ & $+0.019 \pm 0.004$ & $0.136 \pm 0.017$ & 0.025 & 0.023 & 43 \\
$V$ & $17.484 \pm 0.029 $ & $+0.029 \pm 0.003$ & $0.149 \pm 0.015$ & 0.019 & 0.018 & 36 \\
$S$ & $16.611 \pm 0.030 $ & $+0.028 \pm 0.004$ & $0.078 \pm 0.025$ & 0.018 & 0.018 & 32 \\
\hline
\end{tabular}
\label{tab:one}
\end{table*}

\begin{figure*}[t]
\centering
\subfloat[$U-V$ Colour Magnitude Diagram]{\includegraphics[width= 3.35in]{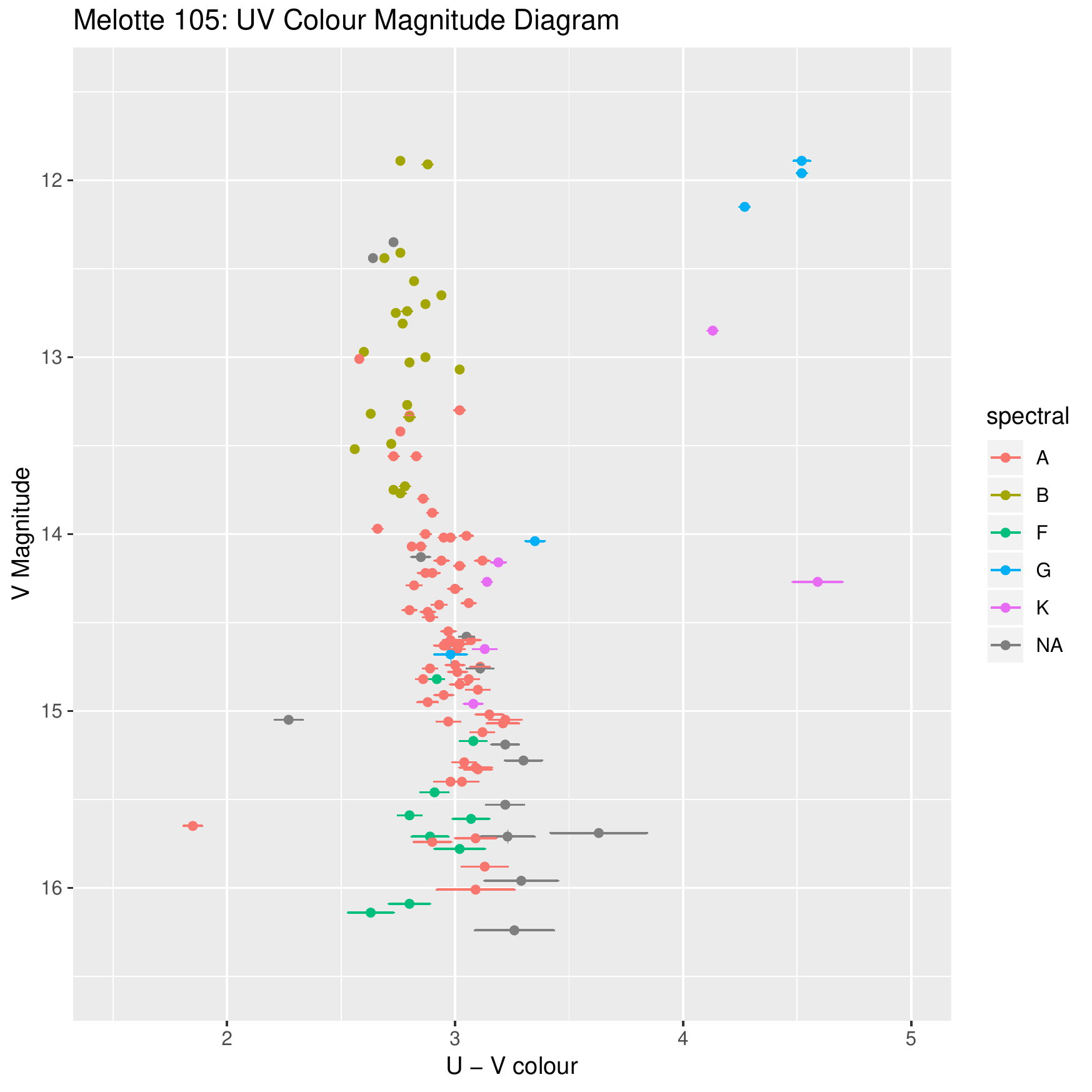}}
\subfloat[$X-V$ Colour Magnitude Diagram]{\includegraphics[width= 3.35in]{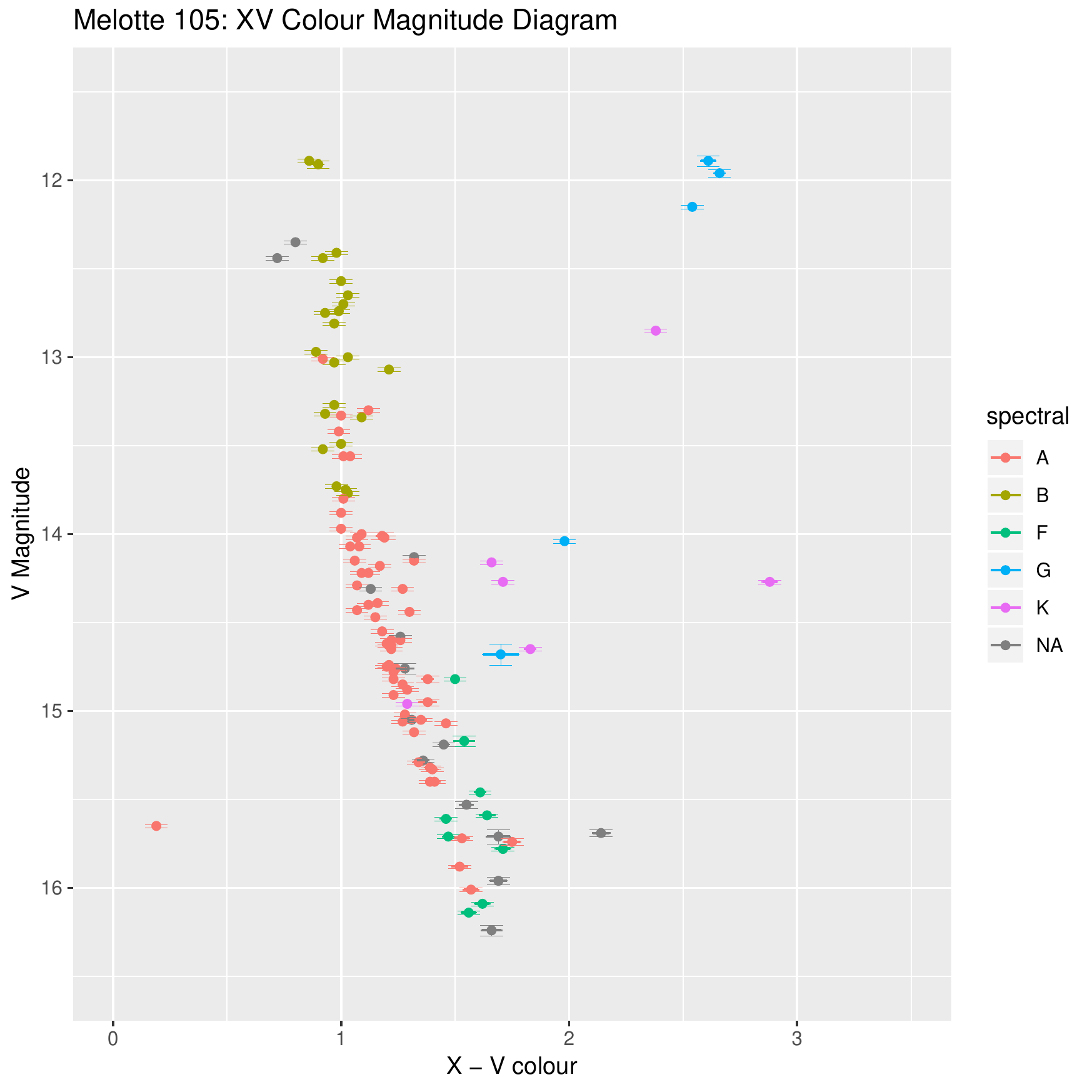}}]]\\
\subfloat[$P-V$ Colour Magnitude Diagram]{\includegraphics[width= 3.35in]{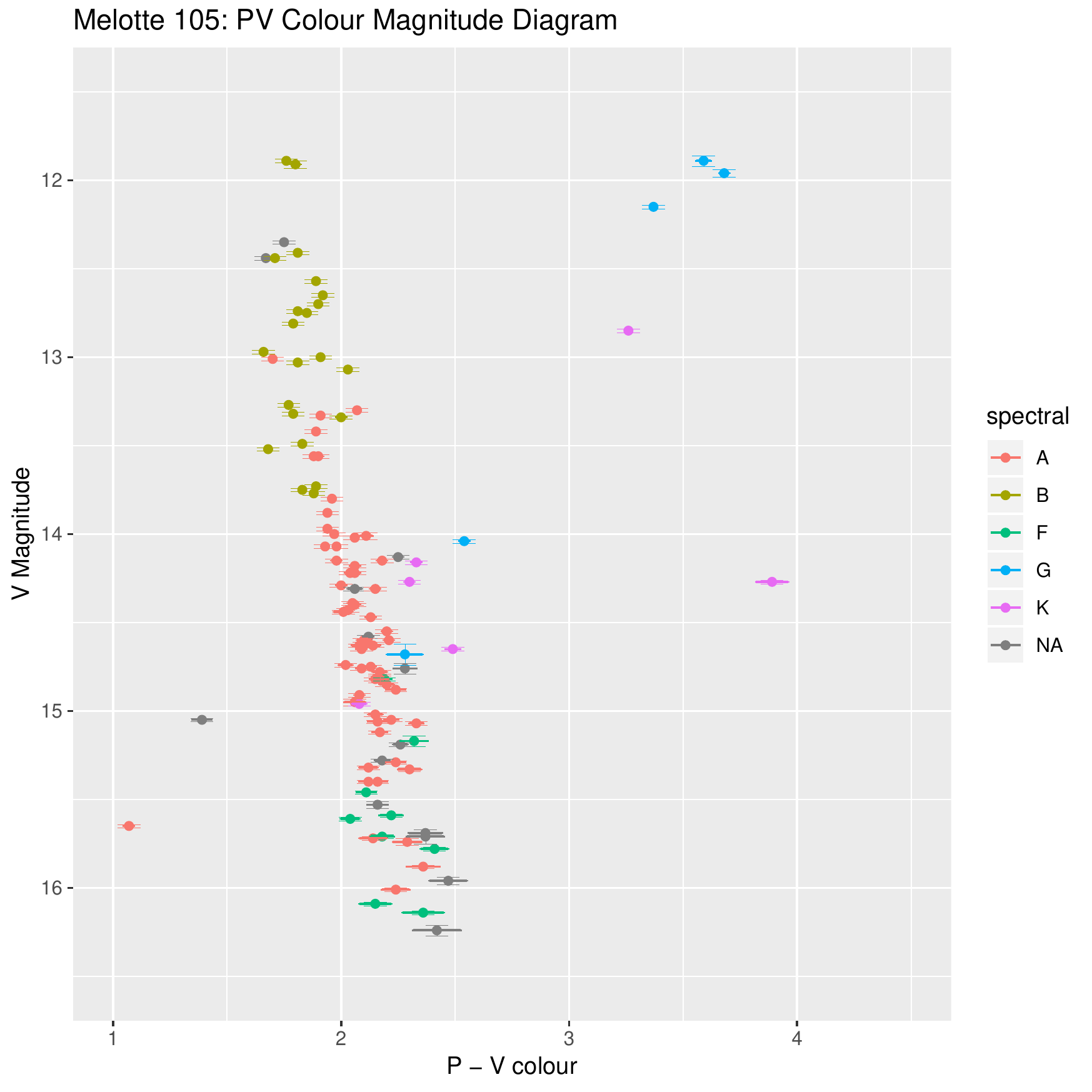}}
\subfloat[$Z-V$ Colour Magnitude Diagram]{\includegraphics[width= 3.35in]{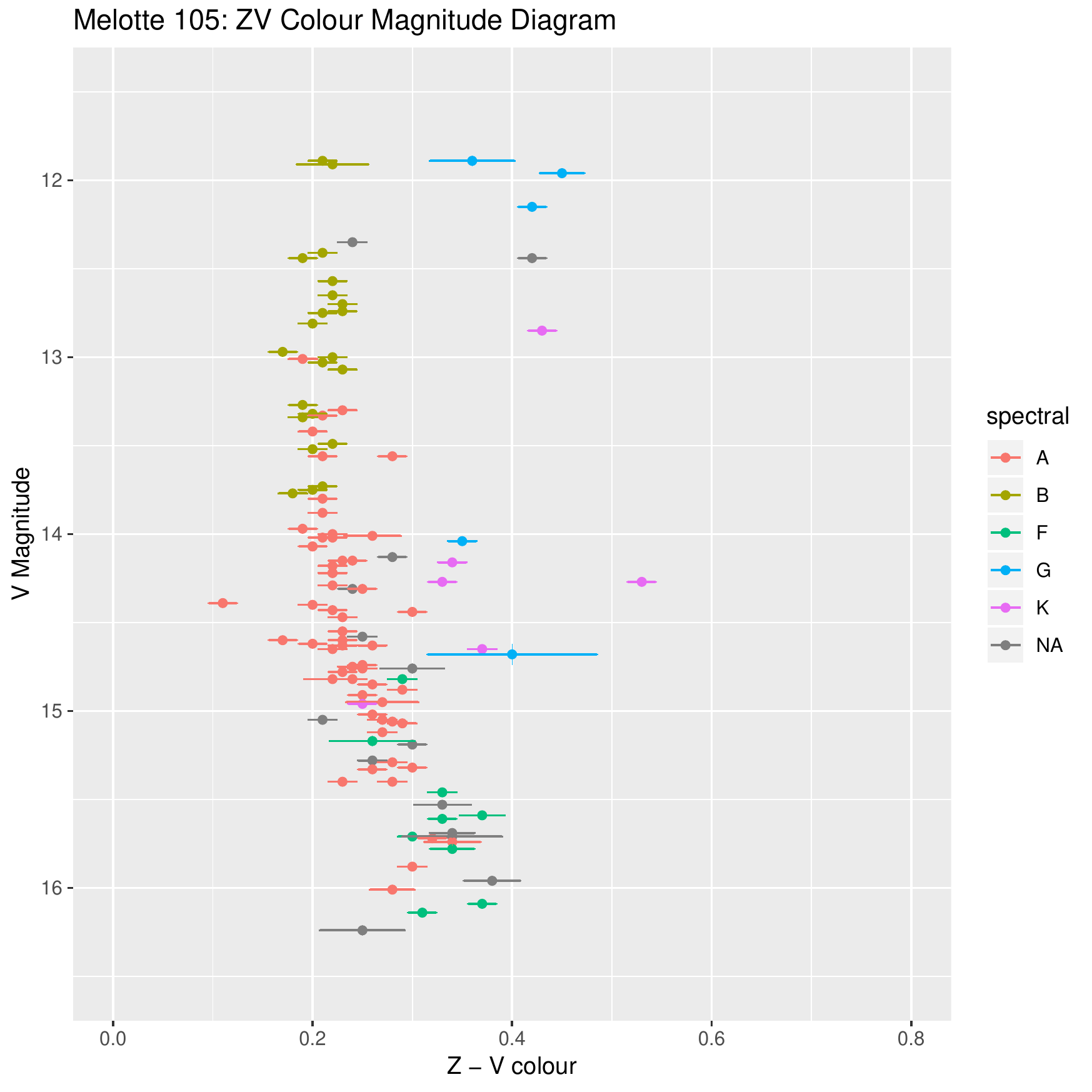}}\\
\caption{{\bf Colour Magnitude Diagrams} 
for Melotte 105, plotting the `standard' colour indices for the Vilnius system.  Data points are coloured according to the derived spectral class. `NA' stands for `not available', indicating those stars where the spectral class could not be derived.}
\label{fig:one}
\end{figure*}


\section{Melotte 105}

Melotte 105 is an open cluster that has not attracted much
attention. It is a compact cluster located in the Galactic plane in
the direction of Carina. In the literature, only a few studies have
produced colour-magnitude diagrams for the cluster. The first was
based on the {\it UBV} photoelectric and photographic photometry of Sher
(1965), who estimated the cluster age via isochrone fits as
$\sim 10^{8}$ years. As part of a search for small photometric
fluctuations in giant and supergiant stars, Frandsen~{\em et al.} (1989)
produced a Johnson {\it BV} CMD based on CCD observations, which extended to
$V \sim 19$ mag. The paper adopted the distance modulus ($V-M_{V}$)=12.7 mag 
and reddening $E(B-V)=0.38$ mag of Sher (1965). Kjeldsen~\& Frandsen (1991) 
re-observed the cluster and derived independent values for these parameters, 
$E(B-V)$ was estimated to be $0.52\pm0.03$ mag, and ($V-M_{V}$) as 
$13.35\pm 0.20$ mag. The authors considered that the difference was due to 
the improved detector used, as well as use of DAOphot (Stetson, 1987) in this 
rather crowded field. However, Kjeldsen~\& Frandsen (1991)
commented that the response of the CCD combined with the $U$ and $B$
filters used by their study differs substantially from the standard
responses and that there were severe problems transforming their data
across to the standard system.

\begin{figure*}[t]
\centering
{\includegraphics[width= 6in]{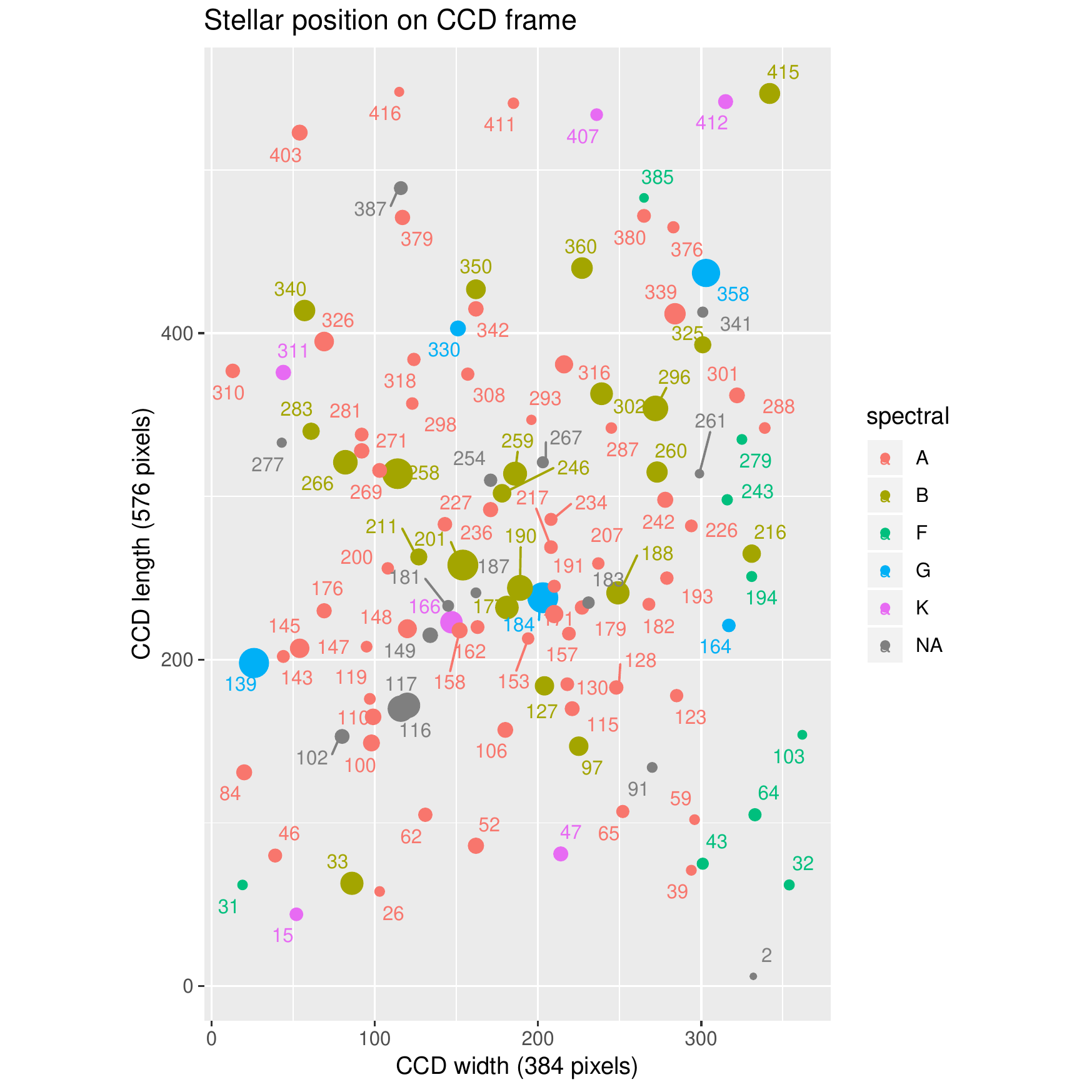}}
\caption{{\bf Positions and Identifications of Stars in Melotte 105:} 
The radius of a point indicating the position of a star is proportional
to the derived $V$ magnitude of that star. The $UPXZVS$ magnitude, 
$x$ and $y$ positions, and identification numbers of the stars plotted 
in this figure are given in Table~\ref{tab:photometry}. A CCD pixel 
corresponds to 0.6 arcseconds at the $f$/7.9 Cassegrain focus of the 
Mclellan telescope}
 \label{fig:two}
\end{figure*}

The cluster was the target of {\em uvby} CCD photometry by Balona \&
Laney (1995). Santos~\& Bica (1993) and Ahumada~{\em et al.} (2000)
performed integrated spectroscopic investigations (deriving $E(B-V)
\sim 0.3$ mag). The reddening, distance and age values derived from all
these studies vary substantially --- by around 50\%, 20\% and 80\%
respectively. Piatti~\& Claria (2001) collected {\it BVI} CCD observations
of the cluster, deriving $E(B-V)=0.42\pm 0.03$ mag, and a ($V-M_{V}$) of 
11.75 mag. Again this is closer to Sher (1965) than to Kjeldsen~\& Frandsen 
(1991).  This should be compared with a $E(B-V)$ of 0.46 mag derived from
Balona~\& Laney (1995) by Piatti~\& Claria (2001), and Sagar~{\em et al.'s} 
(2001) value of 0.52 (no error given) based on {\it UBVRI} CCD photometry, 
which both lean towards the values of Kjeldsen~\& Frandsen (1991). It is 
worth noting that Oliveira~{\em et al.} (2013) report higher values of 
$0.53\pm0.05$ mag for Sagar~{\em et al.'s} (2001) data and $0.50\pm0.02$ mag 
for that of Kjeldsen~\& Frandsen (1991). Dias~{\em et al.} (2012) report 
$0.33\pm0.03$ mag, using 2MASS data.

Given the wide spread and large uncertainties of these studies, it
seemed worthwhile to contribute these old observations of Melotte~105
to the literature, particularly given the novel aspect that the
Vilnius photometric system was used and estimates for spectral types
could be made. {\color{black} The data collected for the current study 
were some of the first CCD imaging using this system 
(see Boyle~{\em et al.}, 1990 a,b, 1992, 1996) and indeed were partly 
collected to test that useful data could be obtained using the system, 
the then available telescopes in New Zealand, and  then recently 
set up data reduction pathway described below.}

\section{Observations}

Observations were made with the 1m McLellan telescope at MJUO (Mount
John University Observatory, New Zealand), using a cryogenically
cooled Thomson TH7882 CDA charge-coupled device. Images were
collected using the Photometrics PM-3000 computer running FORTH (Moore
1974) software with extensive local modifications, and written to half
inch 9 track magnetic tape for transportation back to Victoria
University of Wellington (VUW, New Zealand) for analysis. Images from
these tapes were then converted into the FITS (Wells, Griesen \&
Harten 1981) format from the native Photometrics one, and read into
the Image Reduction and Analysis Facility (IRAF), where subsequent
reduction took place. Details on the data pathway and image processing
facility established at VUW are in Banks (1993). Further details on
the MJUO CCD data acquisition system and its characteristics may be
found in Tobin (1992).


Melotte 105 was observed on February 15th 1994. Seeing was reasonable
for MJUO, with a mean of 3.0 arcsecs for frames of the cluster. All
exposures were 14.35 minutes long. The Vilnius filters were of
different thicknesses, so flats had been attached to bring them all to
the same optical depth (so that refocusing the telescope was not
necessary). The $U$, $P$, $X$, $Y$, $Z$, $V$, and $S$ filters used by 
this study are 9.79, 4.52, 3.26, 6.25, 6.64, 6.79 and 5.91 mm thick respectively.
Schott FK5 was used for the $P$ and $X$ filters, and 8270 for the
remainder, with careful attention paid to not significantly altering
the spectral profiles of the filters. Dow Corning Q2-3067 Optical
Couplant was used to attach suitably thick, pitch polished flats to
the filters.  During the observing run air bubbles increasingly
invaded the optical couplant of the $Y$ filter, making it unsuitable for
imaging. The lack of $Y$ observations meant that the standard
reduction technique for Vilnius data could not be followed for Melotte
105, as it centres around the $Y$ filter. This method is described in
Strai\v zys \& Sviderskien\. e (1972), and Strai\v zys (1992a).

Standard stars were observed in NGC~4755. The IRAF ``Fitparam'' task was 
used to fit equations of the form:
\begin{equation}
\rm M_{o} = M + m_{1} + ( m_{2} \times ( {\it X - V} )) + ( m_{3} \times \text{Airmass} )
\end{equation}
where the subscript `o' indicates the observational magnitudes, m are
the coefficients, and M the `standard' magnitude. The coefficients of
fits and their root mean squares (RMS) may be found in
Table~\ref{tab:one}.  Extinction and instrumental coefficients were
determined by the IRAF ``photcal'' package using a least-squares
solution to all the standard star data (see Harris {\em et al.} 1981).

Aperture corrections add further uncertainties to the transformations,
beyond those given in Table~\ref{tab:one}. The uncertainties in the
aperture corrections for the $U$, $P$, $X$, $Z$, $V$, and $S$ filters 
are 0.029, 0.034, 0.013, 0.014, 0.008, and 0.024 magnitudes, respectively.

Three of Sher's (1965) stars, measured with photoelectric photometry
were in common with the current study.  Agreement was good for the
stars. $V$ magnitudes were {(12.09, 13.00, 14.08) in Sher compared to
 the present study's of {(11.96, 12.95, 14.01)}, while $X-V$ colours 
 were {(2.69, 0.99, 1.16)} compared to {(2.71, 0.92, 1.09)}. The 
 relation given by Forbes (1996) was used to convert the $B-V$ 
 colour to $X-V$.


\section{Results}

{\color{black}
IRAF and its implementation of {\em DAOphot} (Stetson, 1987) were 
used to reduce the CCD images.  Further information on the processing steps
followed may be found in Banks (1993), such as flat fielding and bias removal.
}

\subsection{Photometry}

The cluster has galactic coordinates ($l$, $b$) = (292$^{o}$.90, 
-2$^{o}$.41), and is therefore subject to substantial interstellar 
absorption. Reddening was estimated using a $P-X$, $Z-S$ colour-colour 
diagram as outlined by (Strai\v zys 1992b). Colour excess $E(Z-S)$ is 
estimated as $0.32\pm0.04$ magnitudes, which corresponds to $E(B-V)$ 
being $0.34\pm0.04$. This value is within error of that of Sher (1965), 
but substantially different from the value given by
Kjeldsen \& Frandsen (1991). Santos \& Bica (1993) estimated the
reddening for Melotte 105 using integrated spectra of cluster stars in
the visible and infrared as $E(B-V)=0.30\pm 0.02$ magnitudes,
which is within the formal error of the reddening estimated by the
present study. The reddening was used in conjunction with the
expected luminosity class V sequence to estimate the distance modulus
of the cluster. The true distance modulus $V-M_V$ was found to
be $12.9\pm 0.3$, which is again closer to the value given by
Sher (1965), {\color{black} although with large uncertainty to the extent 
that it also overlaps with Kjeldsen~\& Frandsen (1991)}. 
Paunzen \& Netopil (2006) report a distance modulus of 11.8 (+0.43, -0.53), 
Monteiro {\em et al.} (2010) 11.92 (+0.16, -0.17), and Yadav~\& Sagar (2002) 
12.07 (+0.08, -0.09) magnitudes --- somewhat lower than this study's value.

Given the lack of $Y$ observations, the standard classification technique 
outlined by Strai\v zys (1992a, b) could not be followed. However, 
Strai\v zys (1974) proposed a technique based on the reddening-free energy 
distributions, i.e., using the so-called $Q$ factors. This method could be 
used despite no $Y$ data being available. The $Q$ parameters were calculated 
with respect to the $V-S$ colour index. The parameters $Q_{UVS}$, $Q_{PVS}$, 
$Q_{XVS}$, and $ Q_{ZVS}$ were then plotted against the central wavelength 
of the first filter in each $Q$ factor. Normally $Q_{YVS}$ is included. 
These $Q$ functions have very different shapes for different spectral types 
and luminosity classes.

While this ``poor man's spectroscopy" did not require the $Y$ filter,
the effectiveness of the technique was reduced without it. In
addition, only the brightest stars in the cluster could be classified,
given that the exposures in each filter were of the same length. Hence
the $U$ and $P$ filters had brighter limiting magnitudes, given their
through-puts. The supergiants were classified as late G stars, while
the main sequence extended from $\sim$ B9 through to late A. The
magnitudes, positions, and identification numbers of the classified
stars are given in Table~\ref{tab:photometry}. The table is sorted in
order of increasing $V$ magnitude. The errors given are those calculated
by IRAF {\em DAOphot}. The stars with unusual classifications appear to be
field stars, located on the CMDs well off the main sequence of the
cluster. The major problem with the technique was that the
$ Q_{\lambda}$ functions for most of the spectral range (early to
mid A stars) covered by the main sequence do not change greatly,
especially given the missing $Q_{YVS}$ factor. The $V \times X-V$
colour-magnitude diagram was therefore used to select the appropriate
spectral type, and so the initial colour excess ratios, in the first
iteration of the classification. At least two iterations were used to
produce the results listed in the table. More were required for the
unusual stars. Excesses for later iterations were based on the
spectral type estimated by the immediately previous step.

The classifications are in good agreement with Frandsen {\em et al.}
(1989) who commented that the main sequence for Melotte 105 commenced
at B8. Kjeldsen \& Frandsen (1991) state that Sher (1965) reached down
to the spectral type A5 on the main sequence. The faintest magnitude
tabulated by Sher (1965) is 14.99, which corresponds to A5 V stars in
the current analysis.


{\color{black}
\subsection{GAIA Data}

We compiled the {\it Gaia} DR2 photometric and astrometric catalog of the 
observation results of Melotte 105 determined by Vilnius photometry. 
For this, we matched the star chart of the cluster with {\it Gaia} DR2's 
equatorial coordinates ($\alpha_{2000}$, $\delta_{2000}$), to obtain the 
photometric data ($G$ magnitude and $BP-RP$ colour), trigonometric parallaxes 
($\varpi$), proper motion components ($\mu_{\alpha}\cos \delta$, $\mu_{\delta}$) 
and their errors of the stars located through the region of Melotte 105. 
However, the astrometric data of three of the 116 stars in the catalog is 
not available. We listed photometric and astrometric {\it Gaia} DR2 data of 
the cluster in Table~\ref{tab:gaia}.


\subsection{Structural Parameters of the Cluster}

We constructed the radial density profile (RDP) of Melotte 105 to 
investigate its structural parameters. The central equatorial coordinates 
of the cluster ($\alpha_{2000}$=11$^{\rm h}$ 19$^{\rm m}$ 42$^{\rm s}$, 
$\delta_{2000}$=$-$63$^{o}$ 29$^{'}$00$^{''}$) were taken from SIMBAD 
database\footnote{http://simbad.u-strasbg.fr/simbad/sim-fbasic}. Taking 
into account these coordinates, we calculated the number density of stars 
which are located in 0.5 arcmin-wide concentric circles centered from the 
cluster center to 3.5 arcmin. Then we fitted the RDP with a King (1962) 
model and obtained the cluster's central stellar density $f_0=13.472\pm0.219$ 
star/arcmin$^2$, background stellar density $f_{bg}=0.550\pm0.292$ star/arcmin$^2$, 
and core radius $r_c=1.392\pm0.066$ arcmin. The RDP presented in Fig.~\ref{fig:eps1} 
indicates that most of the stellar density of the cluster is within about 3.5 arcmin.

{\color{black}
\begin{figure*}
\centering
{\includegraphics[width=6in]{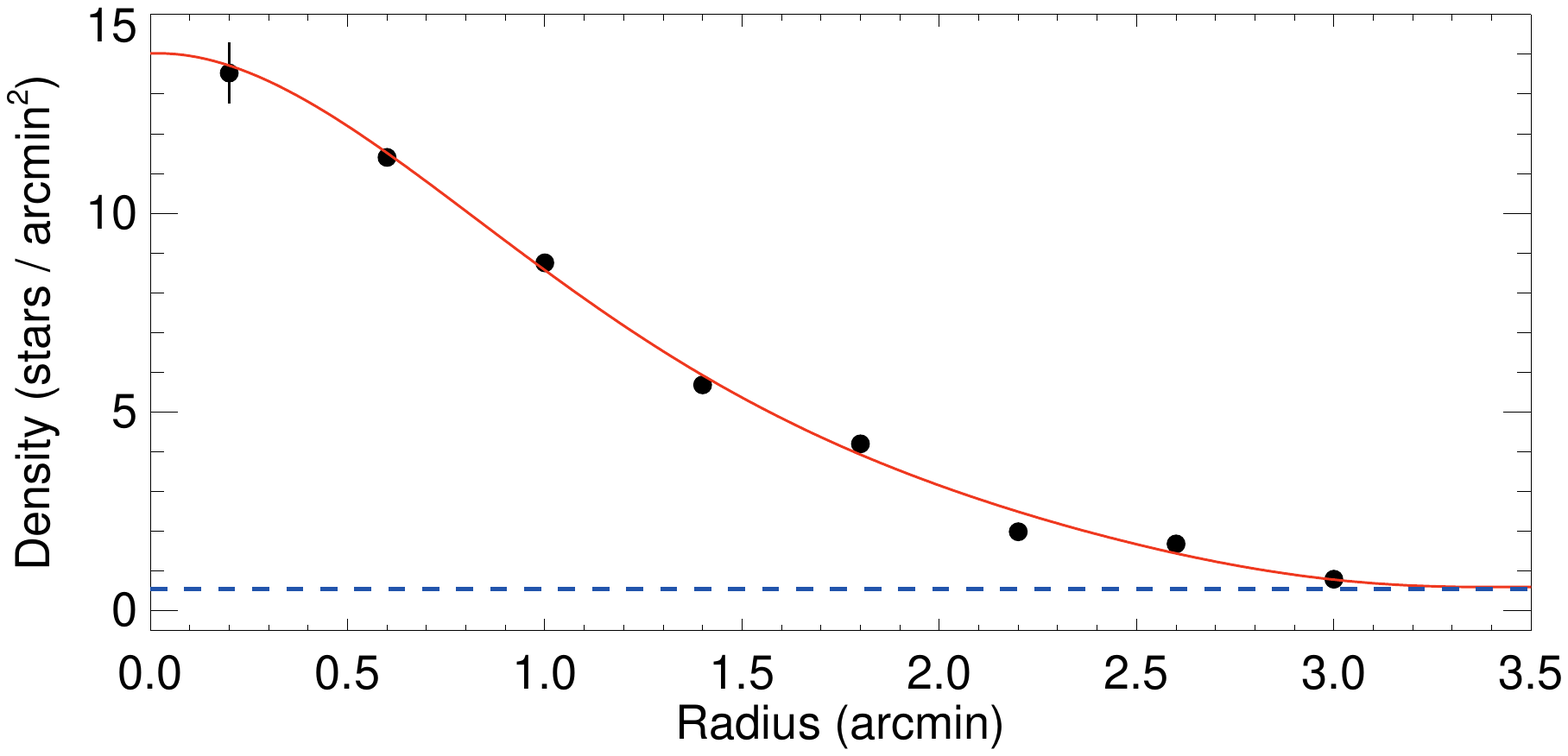}}
\caption{\color{black}
{\bf The radial density profile of the Melotte 105.} The best fit model is shown as the red solid line. Errors were determined as $1/\sqrt{N}$, where $N$ indicates the number of stars used in density estimation. The blue dashed line represents the background stellar density for observational data.}
 \label{fig:eps1}
\end{figure*}

{\color{black}
\begin{figure*}
\centering
{\includegraphics[width=6in]{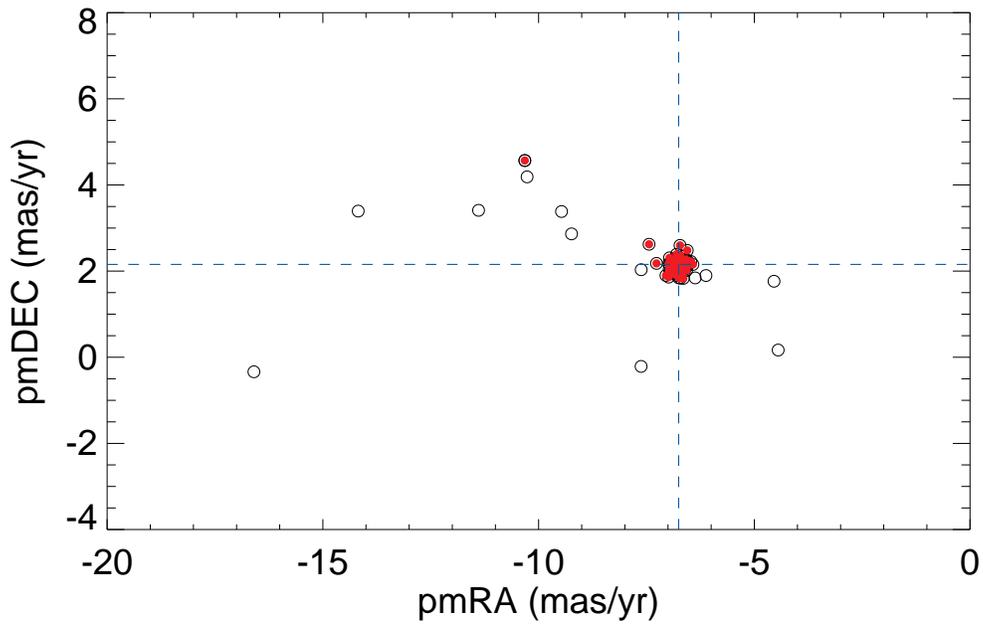}} 
\caption{\color{black} {\bf Membership probability} as calculated by UPMASK. Stars with a probability greater than
$\ge 0.5$ are selected as cluster members by this study.}
 \label{fig:eps2}
\end{figure*}}

{\color{black}
\begin{figure*}
\centering
{\includegraphics[width=6in]{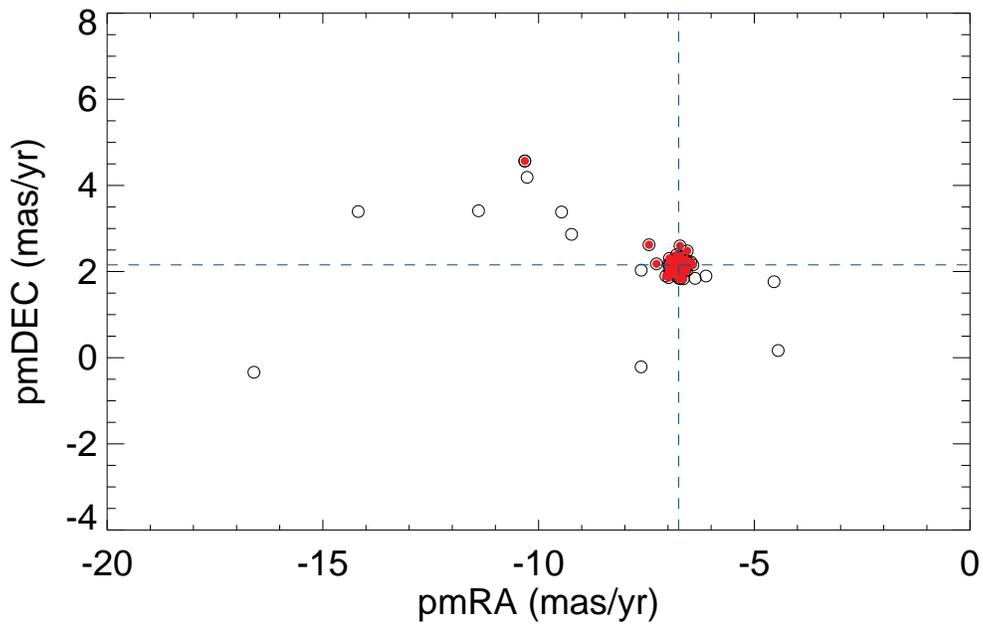}}
\caption{\color{black} 
{\bf The vector point diagram for the observed stars:} red circle denote the most probable member stars, as per
Figure~\ref{fig:eps2}.  The intersection of two blue dashed lines presents the median proper motion values calculated from
the proper motions of these 99 selected stars, reported in the text}
 \label{fig:eps4}
\end{figure*}}

\begin{figure*}
\centering
{\includegraphics[width=6.5in]{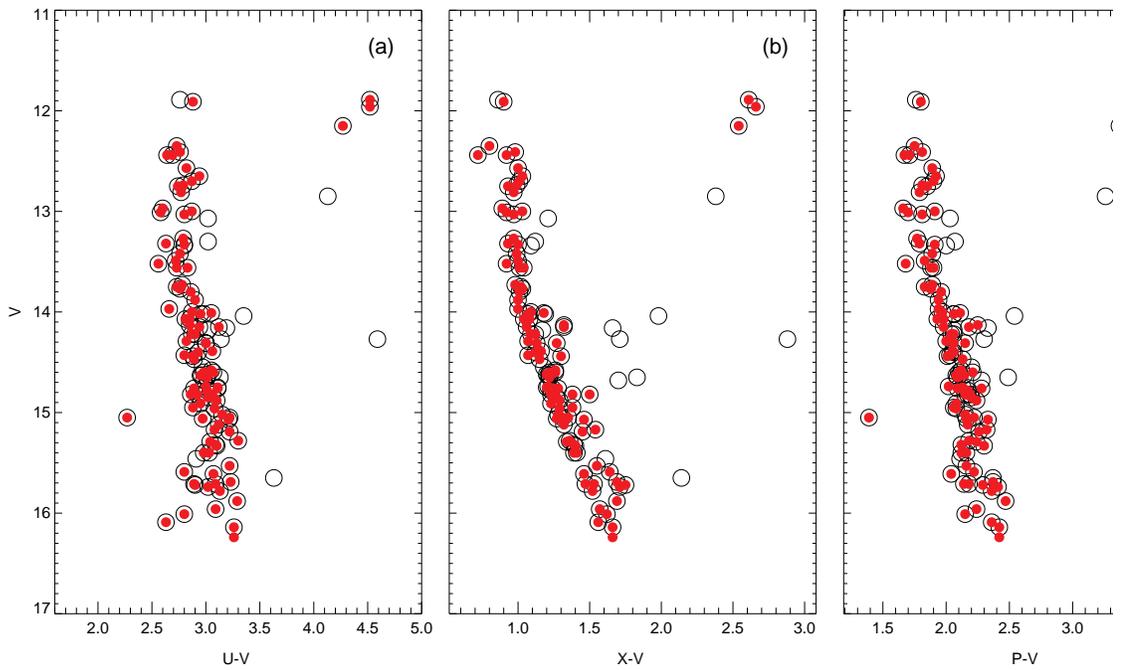}} 
\caption{\color{black} 
{\bf Vilnius Colour Magnitude Diagrams tagged with derived client membership:} red circles denote the most probable member stars, as per
Figure~\ref{fig:eps2}. Subplot (a) is for $U-V$, (b) for $X-V$, (c) for $P-V$, and (d) for $Z-V$.}
 \label{fig:eps3}
\end{figure*}

\begin{figure*}
\centering
{\includegraphics[width=3.25in]{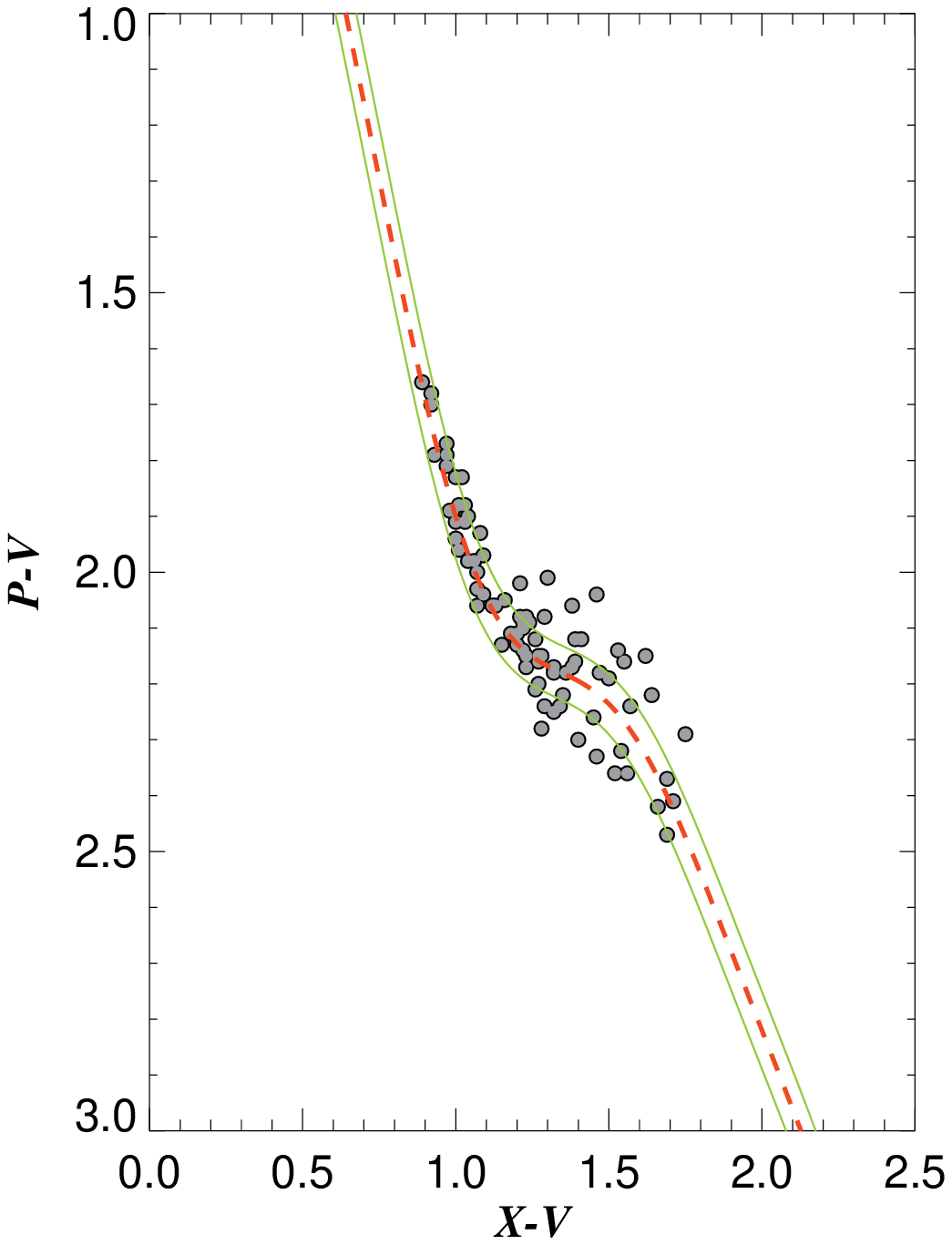}}
{\includegraphics[width=3.25in]{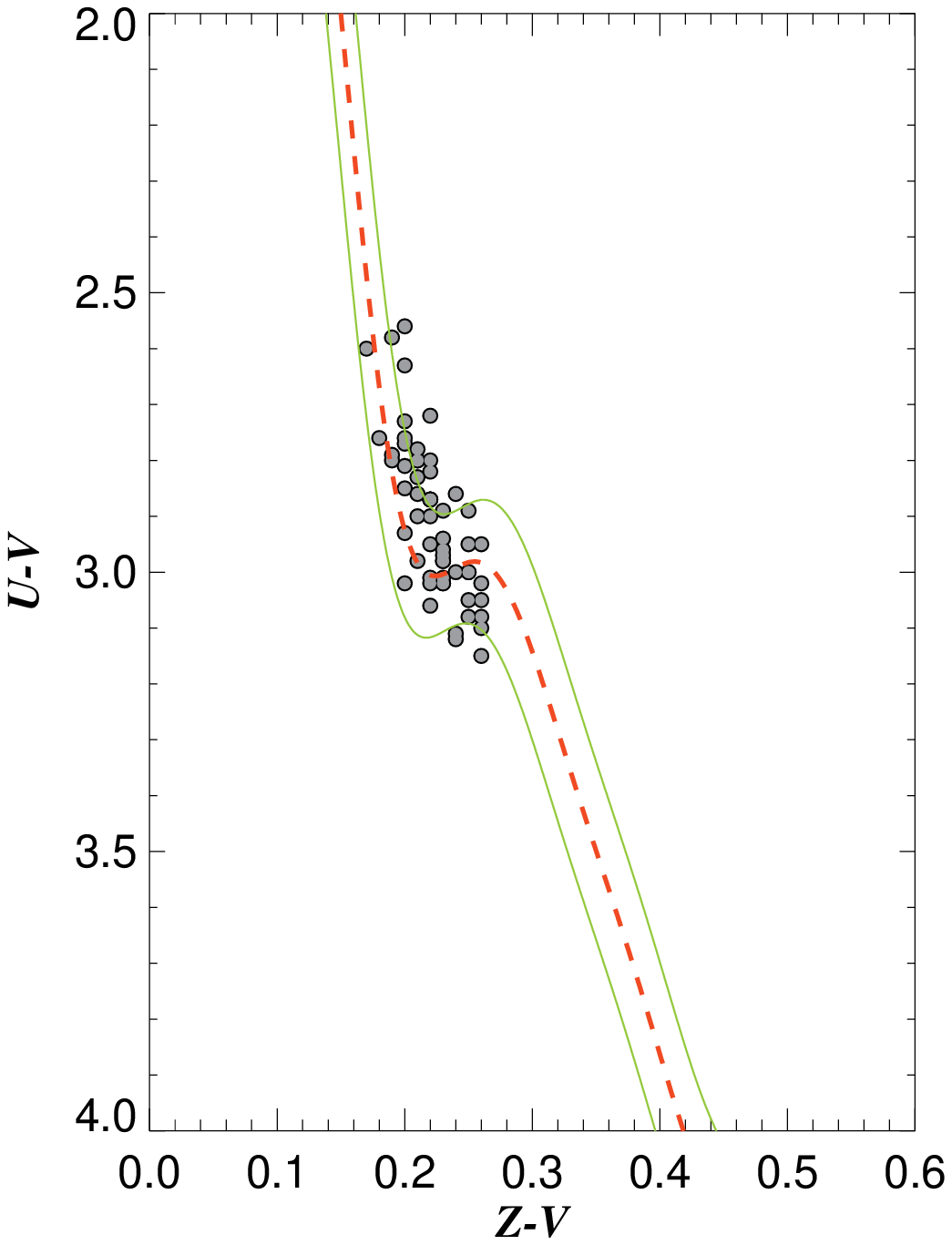}}
\caption{\color{black} 
{\bf Two-colour
diagrams $(P-V)\times (X-V)$ on the left and $(U-V)\times (Z-V)$ on the right.} }
 \label{fig:eps5}
\end{figure*}

\begin{figure*}
\centering
{\includegraphics[width=3in]{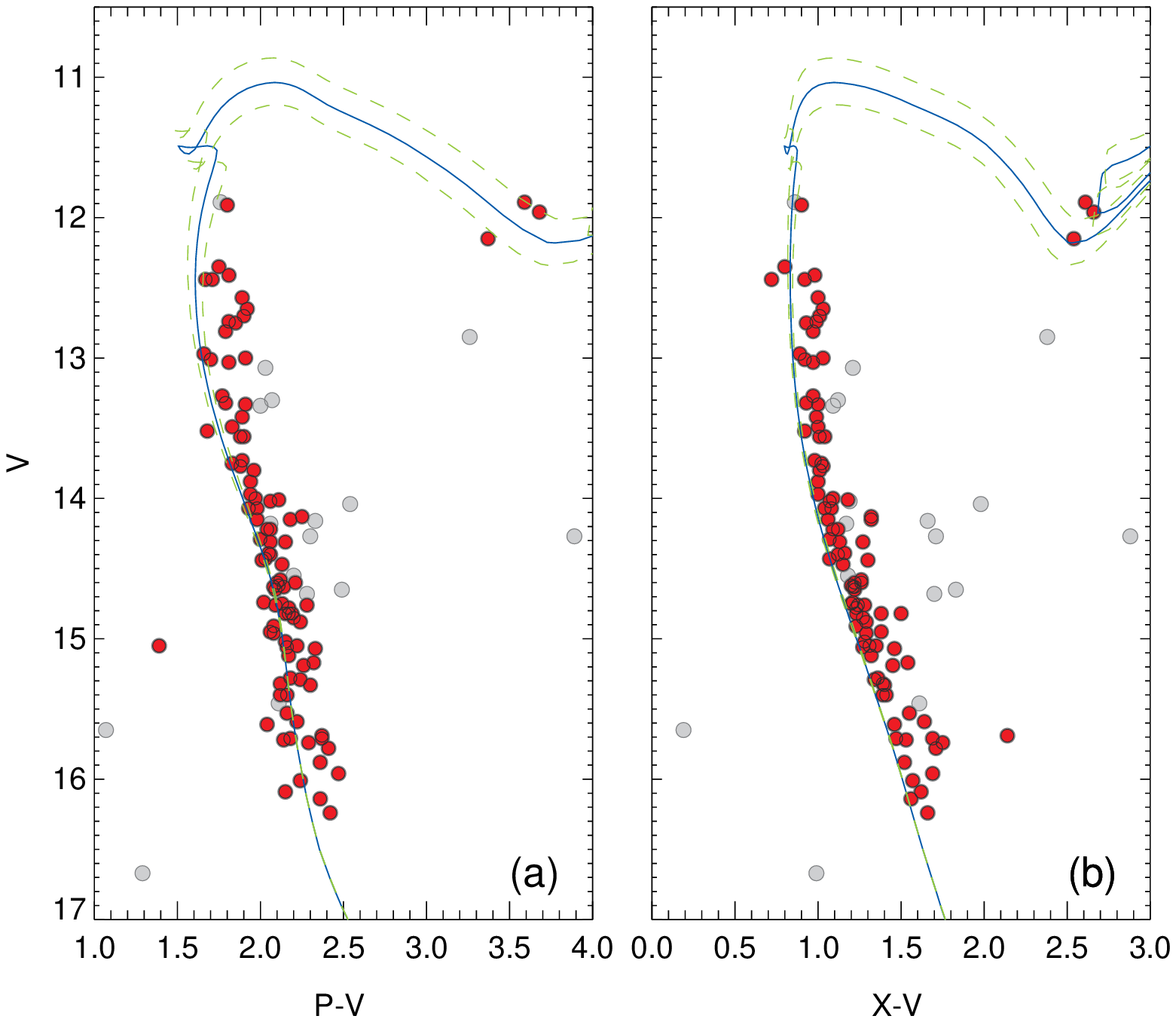}}
{\includegraphics[width=3in]{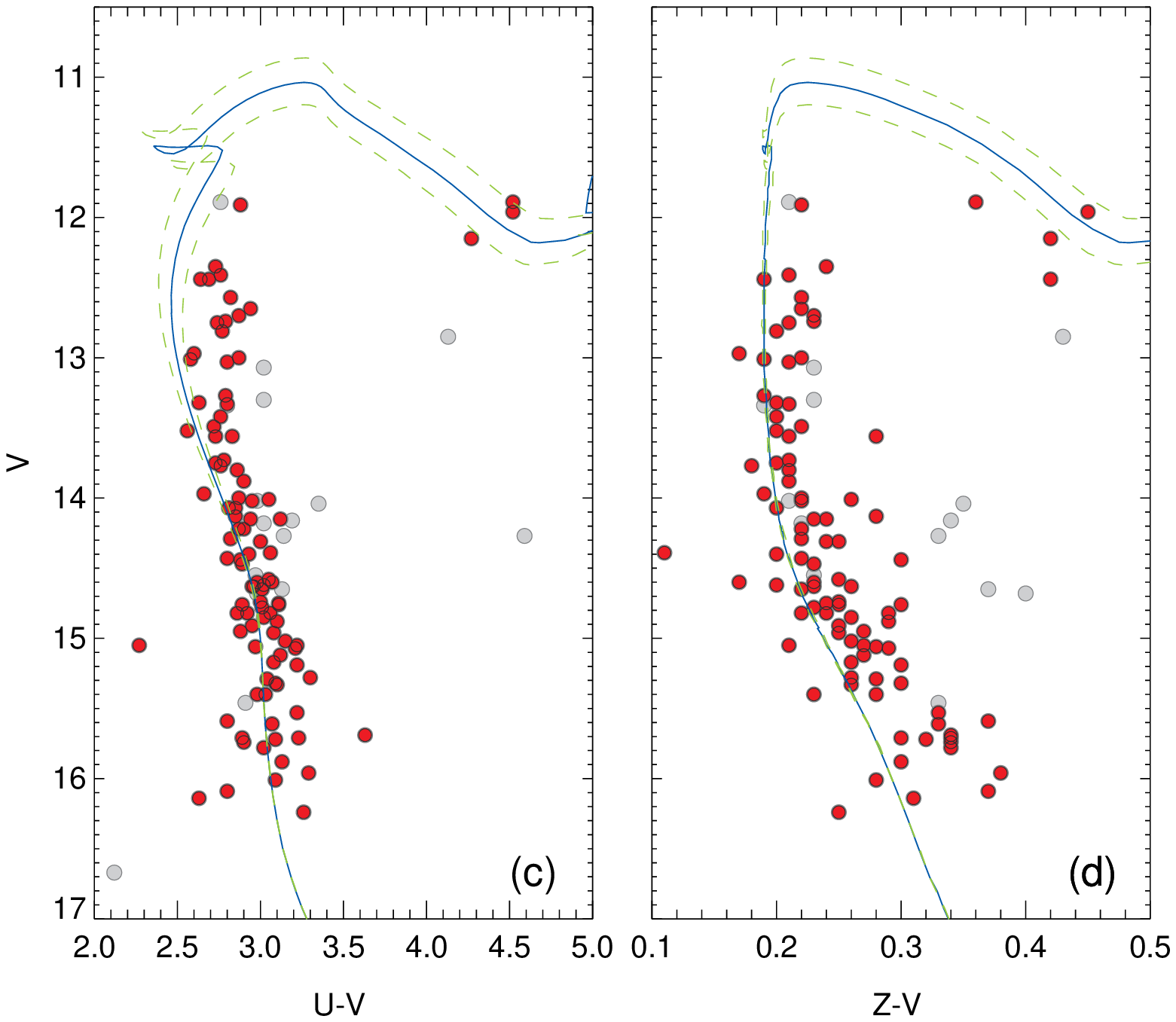}}
\caption{\color{black} 
{\bf Colour Magnitude Diagrams with best fitting {\sc parsec} isochrones.} The parameters used are given in Table~\ref{table:cmd}. The solid lines are the best fits, and the dotted lines the one sigma intervals.}
 \label{fig:eps6}
\end{figure*}

\begin{figure*}
\centering
{\includegraphics[width=6in]{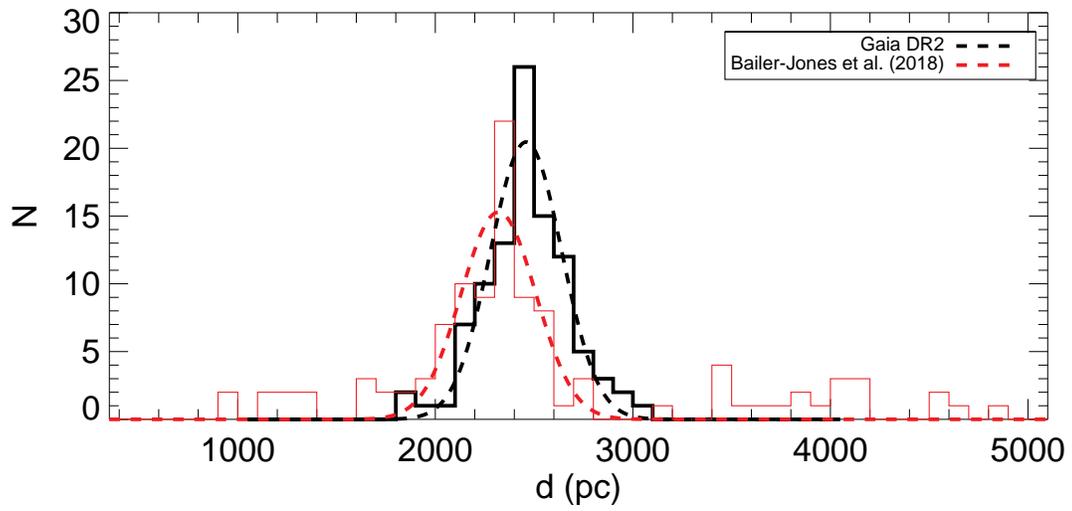}}
\caption{\color{black} 
{\bf Distance histograms} for the Melotte~105 stars with distances calculated by the {\em Gaia} collaboration and Bailer-Jones~{\em et al.} (2018). The dashed lines are the Gaussian fits to these distance histograms.}
 \label{fig:eps7}
\end{figure*}

\begin{figure*}
\centering
{\includegraphics[width=6in]{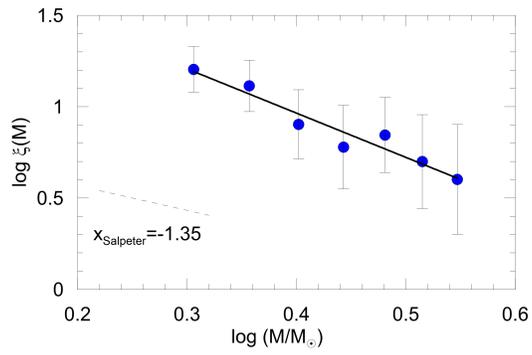}}
\caption{\color{black} 
{\bf Mass function} of Melotte 105 calculated from the most probable main-sequence stars. The solid and dashed lines represent the cluster and Salpeter mass functions respectively.}
 \label{fig:mf}
\end{figure*}

\begin{figure*}
\centering
{\includegraphics[width=6in]{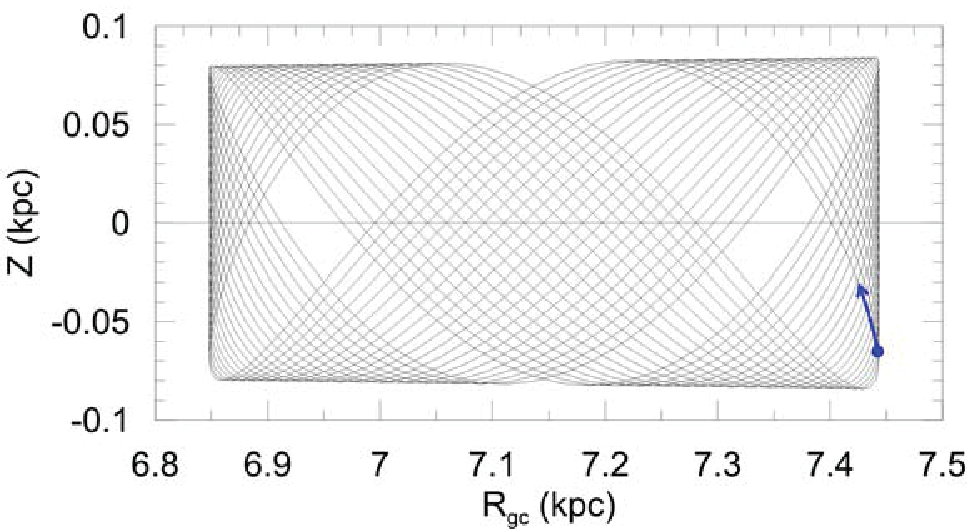}}
\caption{\color{black} 
{\bf The Galactic orbital motions of Melotte 105 in the $R_{gc}\times Z$ plane.} The blue circle and arrow represent the present day position of the cluster and its motion vector, respectively. \label{fig:vel}}

\end{figure*}


\subsection{Cluster membership and CMDs}

To define precise astrophysical parameters of Melotte 105, one should clean the field star contamination from the cluster member stars. In most cases colour-magnitude diagrams (CMDs) of clusters are insufficient to separate field stars from the cluster stars. For those cases, there are some statistical methods that take into account the proper motion values of the stars (Ak ~{\em et al.} 2016; Krone-Martins~\& Moitinho, 2014; Javakhishvili~{\em et al.},  2006; Balaguer-Nunez~{\em et al.}, 1998) to calculate membership probabilities of stars, determining whether the stars are components of the  the cluster or not.  

In this study, we used $UPXZVS$ Vilnius magnitude, colours and {\it Gaia} DR2 astrometric data ($\mu_{\alpha}\cos \delta$, $\mu_{\delta}$ and $\varpi$) and utilized a statistical method called UPMASK (Unsupervised Photometric Membership Assignment in Stellar Clusters, Krone-Martins~\& Moitinho, 2014) in order to calculate membership probabilities. The UPMASK method relies on minimal physical assumptions about high density stellar regions and is suitable for clusters 2-2.5 kpc from the Sun. We applied it to the five-dimensional astrometric space based on equatorial coordinates, proper motions and parallaxes of stars and four-dimensional photometric space ($V\times P-V$, $V\times X-V$, $V\times U-V$, $V\times Z-V$). We calculated the membership probabilities ($P$, not to be confused with the Vilnius filter) of the 116 stars located through Melotte 105 and showed their distribution on membership probability histogram (Fig.~\ref{fig:eps2}). As shown in this figure, the membership probabilities of 99 of 116 stars are greater than $P=0.5$, so these stars are adopted as most probable members of the Melotte 105 and used in further analysis. In fact, when the vector point diagram (VPD) is examined (Fig.~\ref{fig:eps4}), it is seen that stars with $P\geq0.5$ are concentrated in a certain proper motion space. The red circles denote the most probable member stars $P\geq0.5$ while open circles represent the low probable stars ($P<0.5$) located through the cluster region. Intersection of the blue dashed lines represents the median proper motion values that are calculated from proper motions of 99 most probable cluster member stars. The median values of proper motion components are as follow: ($\langle \mu_{\alpha}\cos \delta \rangle=-6.753\pm0.041$, $\langle\mu_{\delta}\rangle=2.156\pm0.040$ mas/yr). Next, we constructed CMDs of the Melotte 105 for four photometric colours ($V\times U-V$, $V\times X-V$, $V\times P-V$, $V\times Z-V$) and marked the stars with $P\geq0.5$ on these diagrams (see Fig. 6.) When we examine the figure by eye, it is seen that the main-sequence, turn-off point and giant stars can be distinguished. The turn-off points of CMDs contain stars whose apparent $V$ magnitudes are between $\sim$11.5 and 13 mag, while evolved stars are bright ($V\sim 12$ mag) and they are in the red colour region (i.e. $X-V\sim 2.5$ mag) of the diagrams.


\subsection{Colour Excess Determination:} 


To calculate the colour excesses towards the cluster, we used the 84 main-sequence stars whose membership probabilities are $P\ge 0.5$ and with apparent magnitudes $V\ge 12$. The de-reddened ZAMS relations for Vilnius photometry are taken from Strai\v zys  (1992a). The two colour diagrams  (TCDs) $ (P-V) \times (X-V)$ and $(U-V) \times (Z-V)$ were used to estimate the reddening (see Figure~\ref{fig:eps5}). The selective absorption coefficients\footnote{$\frac{A_U}{A_V} = 1.621$, $\frac{A_P}{A_V} = 1.511$, $\frac{A_X}{A_V} = 1.419$, $\frac{A_Z}{A_V} = 1.082$, $\frac{A_{VIL}}{A_V} = 1.004$, and $A_{V} = 3.1 \times E(B-V)$.} for Vilnius photometry were
taken from Munari~\& Fiorucci (2003) and used in the calculation of the reddening vectors. Based on these, the slopes of the reddening vectors were calculated as $\frac{E(P-V)}{E(X-V)} = 1.223$ and $\frac{E(U-V)}{E(Z-V)} = 7.901$ and used to adjust the position of the ZAMS on the TCDs.  Minimum $\chi^2$ fitting derived the following estimates of the colour excesses and their one-sigma intervals: $E(P-V)=0.637 \pm 0.022$, $E(X-V)=0.521 \pm 0.018$, $E(U-V)=0.877 \pm 0.055$, and $E(Z-V)=0.111 \pm 0.007$. Using the relation $E(X-V)=1.285 \times E(B-V)$, we obtained $E(B-V)=0.403 \pm 0.020$ magnitudes, which is larger than the estimate above of $E(B-V)=0.34 \pm 0.04$ mag although the two sigmas intervals overlap. In passing we note that the $Q$-factor method used above to calculate spectral type estimates is a reddening free technique (see Strai\v zys, 1974, who describes it as a method ``insensitive to interstellar reddening''), and so these estimates do not need to be recalculated based on this refined reddening estimate.

{\color{black}
\begin{table*}[tb]
\centering
\caption{\color{black}{\bf Parameters derived in the {\em Gaia} analysis:} 
where `BJ' stands for the distance (in parsecs) calculated from the 
{\em Gaia} DR2 trigonometric parallaxes and stellar distances given by 
Bailer-Jones~{\em et al.} (2018), `Isochrones' is based on the isochrone 
fitting described by the paper, and `{\em Gaia}' on the DR2 distances of 
the 99 most probable cluster members from this paper. [Fe/H] is the stellar 
metallicity defined using the total iron content compared to the Solar, 
while $Z$ is overall mass fraction for metals (i.e., excluding H and He).}
\begin{tabular}{|c|c|c|c|c|c|c|}
\hline
Distance Modulus (mag) & Z & [Fe/H] (dex) & Age (Myr) & Isochrones (pc) & {\em Gaia} (pc) & BJ (pc) \\
\hline
$12.85 \pm 0.07 $ & 
0.025 &
0.24 &
$240 \pm 25 $ &
$2078 \pm 78$ &
 $2460 \pm 180$ & 
 $2320 \pm 190$\\
\hline
\end{tabular}
\label{table:cmd}
\end{table*}

\subsection{Age Determination} 

The age of Melotte 105 was determined by comparing the PARSEC isochrones 
(Bressan~{\em }et al.}, 2012; Tang~{\em et al.}, 2014; Chen~{\em et al.}, 
2014) for $Z=0.025$ with the observed CMDs (see Figure~\ref{fig:eps6}). 
Metal abundance calibrations are sensitive to the $Y$ band, which is 
unfortunately lacking in the Vilnius data set obtained by this study. 
We were therefore not able to determine the metal abundance of the cluster 
using the photometric metal abundance calibrations. Hence the metallicity 
of the cluster was not estimated via an independent method. For this 
reason during age estimation of the Melotte 105, the distance modulus and 
metallicity are derived simultaneously excluding reddening. We used PARSEC 
isochrones with different ages and metallicities to obtain the best fit results to
$V \times  (P-V)$, $V \times  (X-V)$, $V \times (U-V)$ and $V\times (Z-V)$ CMDs. 
To do this we took into account the standard selective absorption coefficient as 
$R_{V}=3.11$ for Vilnius photometric system (Munari~\& Fiorucci, 2003). The 
uncertainties in age ($\Delta t=25$ Myr) and distance ($\Delta \mu = 0.07$ mag) 
are suitable for the distribution of the most cluster member stars. The $Z$ metal 
abundance was transformed to [Fe/H] using equations given by Bovy\footnote{https://github.com/jobovy/isodist/blob/master/isodist/\\Isochrone.py} 
who analytically obtained them using PARSEC isochrones (see also Yontan~{\em et al.}, 
2019; Bostanc\i~{\em et al.}, 2018; Bilir {\em et al.}, 2016) and calculated the 
metallicity as [Fe/H]=0.24 dex. We represented the CMDs with the best fit PARSEC 
isochrones in Figure~\ref{fig:eps6} and the results were listed in 
Table~\ref{table:cmd}. Melotte 105 is estimated to be $240\pm25$ Myr old. The 
cluster distance derived from the main-sequence fitting is 
$2078\pm78$ pc. We compare this with Bailer-Jones~{\em et al.} (2018) who inferred 
distances to essentially all 1.33 billion stars with parallaxes published in the 
{\it Gaia} DR2 and our estimate using the {\it Gaia} distances for the 99 most 
probable cluster members together with the relation distance $d=1000/\varpi$. Gaussian 
fitting to both these data sets (see Figure~\ref{fig:eps7}) led to the estimates 
given for these studies in Table~\ref{table:cmd}. The estimate from 
Bailer-Jones~{\em et al.} (2018) is more compatible with the estimate based on 
the isochrones fitting.


\subsection{Mass function} 

We obtained a high degree polynomial function between the absolute magnitudes 
and masses of the main-sequence stars based on the best fitting {\sc parsec} 
isochrones. $V$ apparent magnitudes were converted to absolute magnitudes for 
the main-sequence cluster member stars ($P\ge 0.5$) using the derived distance 
modulus of the cluster. The member star theoretical masses were calculated from 
the derived absolute magnitude-mass relation. We obtained the mass function slope 
from the relation $\log \xi(M)=-(1+X) \times \log{M}+C$ where $C$ is a constant. 
Figure~\ref{fig:mf} shows the mass function for Melotte 105. The slope of this 
mass function was obtained from 59 stars whose masses range 1.90 to 3.52 
$M/M_{\odot}$. The mass function slope variable $X$ is  $-1.42\pm0.29$, in very 
good agreement with Salpeter's (1955) $X=-1.35$ value.


\subsection{Galactic Orbit}

The galactic orbit of the cluster was calculated using the potential functions 
that defined in {\sc galpy}, the Galactic dynamics library 
(Bovy, 2015)\footnote{See also https://galpy.readthedocs.io/en/v1.5.0/}.
The calculation assumed an axisymmetric potential for the Milky Way galaxy, 
following {\sc MWPotential2014} (Bovy, 2015). A circular velocity and the 
galactocentric distance of the Sun were assumed as $V_{rot}=220$ km/s and 
$R_{gc}=8$ kpc (Majewski, 1993) respectively.

The {\sc MWPotential2014} code (Bovy, 2015) was used to determine the space 
velocity components and galactic orbital parameters of the cluster. Calculations 
require equatorial coordinates, mean distance, proper motion components, and 
radial velocity of Melotte 105. As mentioned in previous steps in this section, 
the distance of the cluster is $d=2078 \pm 78$ pc, and the median proper motion 
values were obtained as $\langle \mu_{\alpha}\cos\delta \rangle=-6.753\pm0.041$ 
and $\langle \mu_{\delta} \rangle =2.156\pm0.040 $ mas per year. Among the most 
probable cluster member stars, there are four stars whose radial velocities were 
available in the literature. The radial velocities of three of the four stars 
were present in both the Mermilliod~{\em et al.} (2008) 
(ID 117: $V_R=-4.32\pm0.37$} km/s,  ID 139: $V_R=0.54\pm0.18$ km/s, \& ID 184: 
$V_R=-0.15\pm0.31$ km/s) and the {\em Gaia} DR2 catalogue (ID 117: 
$V_R=-3.56\pm0.84$ km/s, ID 139: $V_R=9.22\pm9.16$ km/s, \& 
ID 184: $V_R=0.63\pm1.66$ km/s), while the radial velocity of one star was 
found only in {\em Gaia} DR2 (ID 358: $V_R=-0.80\pm1.91$ km/s). Where 
measurements were available in both catalogues, the measurement with the small 
radial velocity error was preferred by this analysis. A weighted average was 
taken, leading to the cluster mean radial velocity being taken as 
$-1.11\pm0.14$ km/s. Kinematic and dynamic calculations were analyzed with 
2 Myr steps over a 3 Gyr integration time. 

The space velocity components of the cluster were calculated as $(U, V, W)$=
($-64.75\pm3.34$, $-26.01\pm5.69$, $-3.12\pm0.41$) km/s. To obtain the space 
velocity more precisely, the first-order galactic differential rotation 
correction was taken into account (Mihalas \& Binney, 1981), with $-52.02$ 
and $1.94$ km/s differential rotation corrections applied to $U$ and $V$ 
space velocity components respectively. The $W$ velocity is not affected in 
this first-order approximation. As for the local standard of rest correction 
values from Co{\c s}kuno{\v g}lu~{\em et al.} (2011) of ($8.83\pm0.24$, 
$14.19\pm0.34$, $6.57\pm0.21$) km/s were used leading to corrected values of 
the space velocity components of the cluster of $(U, V, W)$ = ($-3.90\pm3.34$, 
$-13.76\pm5.69$, $+3.45\pm0.41$) km/s. 
 
Melotte 105's perigalactic and apogalactic distances were obtained as 
$R_{p} = 6.85$ and $R_{a} = 7.44$ kpc, respectively. The maximum vertical 
distance from the Galactic plane was calculated as $Z_{max}=84$ pc and the 
eccentricity of the orbit was determined as $e=0.042$. 
 
A representation of the Galactic orbit for Melotte 105 on the $R_{gc}\times Z$ 
plane is shown in Figure~\ref{fig:vel}. Kinematic and dynamic calculations for 
the cluster support that the cluster was formed in a metal-rich environment 
within the Sun circle.  Wu~{\em et al.} (2009) gave the cluster distance 
$d= 2208\pm442$ pc and the radial velocity $V_{R}=0.4\pm0.2 $ km/s. These 
values are similar to those of the current study, but the proper motion 
components $\langle \mu_{\alpha}\cos\delta \rangle=-4.37\pm0.82$ and 
$\langle \mu_{\delta} \rangle=-4.07\pm0.96$ mas/yr are different, leading to 
their orbital parameter estimates being inconsistent with this study. We 
believe the use of the {\em Gaia} DR2 precise astrometric data is an improvement.

}}

\section{Conclusions} 

{\color{black} This paper suggests} the need for further study of
{\color{black} Melotte~105}, although there does appear to be a building convergence towards similar estimates in the literature. {\color{black} Melotte~105 is a suitably populous cluster for a later, careful study investigating completeness and adjusting the mass function suitably} to estimate the cluster's initial mass function (similar to Mateo (1987), 
Sagar \& Richtler (1991), and Banks~{\em et al.} (1995)).
{\color{black} In the near future the Melotte 105 cluster could be studied via 
high resolution and high S/N spectroscopic observations. The cluster member stars then
could be determined using radial velocity analysis, together with the model 
atmosphere parameters of the member stars and the mean metal abundance 
being obtained precisely.}

The {\color{black} current study} is somewhat unusual in that a relatively uncommon photometric system was employed, demonstrating the potential of the Vilnius system to the study of populous clusters.  Longer exposure times, a larger telescope, or a combination of both would result in smaller photometric errors in both the cluster and standard star photometry. Together with inclusion of the $Y$ filter, this could lead to a definitive study of the cluster and a final resolution of the discrepancies in the literature. 


\section{Acknowledgements}

TB is grateful for generous time allocations at Mount John
University Observatory, to the Vilnius Observatory for supplying the
filter set, the New Zealand Lottery Board for financing the purchase
of the set, to Acorn New Zealand for the loan of a R260 computer on
which some of this work was performed, and to the Foundation for Research,
Science and Technology for partial funding of this project in
conjunction with the VUW Internal Research Grant Committee. He acknowledges partial support during this study by the inaugural R.H.T. Bates Postgraduate
Scholarship.

{\color{black} We thank the anonymous referee for their helpful comments and advice which improved this paper.}

This work has made use of data from the European Space Agency (ESA) mission
{\it Gaia} (\url{https://www.cosmos.esa.int/gaia}), processed by the {\it Gaia} Data Processing and Analysis Consortium (DPAC,
\url{https://www.cosmos.esa.int/web/gaia/dpac/consortium}). 

Funding for the DPAC has been provided by national institutions, in particular the institutions participating in the {\it Gaia} Multilateral Agreement.

IRAF was distributed by the National Optical Astronomy Observatories,
which were operated by the Association of Universities for Research in
Astronomy, Inc., under contract with the National Science Foundation.


\clearpage
\onecolumn

\begin{longtable}{llllllllllllllll}
\caption{Derived photometric values for Melotte 105. Each of the Vilnius magnitudes is given (U, P, X, Z, V, S) along with the
photometric errors (dU, dP, dX, dZ, dV, dS), position on the CCD frame (x and y), identification number assigned by this study (ID), and 
the derived classification (Class).  \label{tab:photometry}}\\
 \hline
 \textbf{U} & \textbf{dU} & \textbf{P} & \textbf{dP} & \textbf{X} & \textbf{dX} & \textbf{Z} & \textbf{dZ} &
 \textbf{V} & \textbf{dV} & \textbf{S} & \textbf{dS} & \textbf{x} & \textbf{y} & \textbf{ID} & \textbf{Class} \\
\hline
\endfirsthead
\multicolumn{16}{c}%
{\tablename\ \thetable\ -- \textit{Continued from previous page}} \\
\hline
 \textbf{U} & \textbf{dU} & \textbf{P} & \textbf{dP} & \textbf{X} & \textbf{dX} & \textbf{Z} & \textbf{dZ} &
 \textbf{V} & \textbf{dV} & \textbf{S} & \textbf{dS} & \textbf{x} & \textbf{y} & \textbf{ID} & \textbf{Class} \\
 \hline
 \endhead
16.41 & 0.02 & 15.48 & 0.01 & 14.50 & 0.01 & 12.25 & 0.03 & 11.89 & 0.03 & 10.88 & 0.04 & 203 & 238 & 184 & G8I\\
14.65 & 0.01 & 13.65 & 0.01 & 12.75 & 0.01 & 12.10 & 0.01 & 11.89 & 0.01 & 11.40 & 0.01 & 154 & 258 & 201 & B8III\\
14.79 & 0.01 & 13.71 & 0.01 & 12.81 & 0.01 & 12.13 & 0.03 & 11.91 & 0.02 & 11.38 & 0.01 & 114 & 314 & 258 & B8III\\
16.48 & 0.01 & 15.64 & 0.01 & 14.62 & 0.01 & 12.41 & 0.01 & 11.96 & 0.02 & 11.06 & 0.02 & 26 & 198 & 139 & G8I\\
16.42 & 0.02 & 15.52 & 0.01 & 14.69 & 0.01 & 12.57 & 0.01 & 12.15 & 0.01 & 11.23 & 0.01 & 303 & 437 & 358 & G5I\\
15.08 & 0.01 & 14.10 & 0.01 & 13.15 & 0.01 & 12.59 & 0.01 & 12.35 & 0.01 & 12.16 & 0.05 & 116 & 170 & 116 & NA\\
15.17 & 0.01 & 14.22 & 0.01 & 13.39 & 0.01 & 12.62 & 0.01 & 12.41 & 0.01 & 12.04 & 0.01 & 189 & 244 & 190 & B9V\\
15.08 & 0.01 & 14.11 & 0.01 & 13.16 & 0.01 & 12.86 & 0.01 & 12.44 & 0.01 & 11.52 & 0.03 & 120 & 172 & 117 & NA\\
15.13 & 0.01 & 14.15 & 0.01 & 13.36 & 0.01 & 12.63 & 0.01 & 12.44 & 0.01 & 12.06 & 0.01 & 272 & 354 & 296 & B8V\\
15.39 & 0.01 & 14.46 & 0.01 & 13.57 & 0.01 & 12.79 & 0.01 & 12.57 & 0.01 & 12.17 & 0.01 & 82 & 321 & 266 & B9V\\
15.59 & 0.01 & 14.57 & 0.01 & 13.68 & 0.01 & 12.87 & 0.01 & 12.65 & 0.01 & 12.25 & 0.01 & 186 & 314 & 259 & B9V\\
15.57 & 0.01 & 14.60 & 0.01 & 13.71 & 0.01 & 12.93 & 0.01 & 12.70 & 0.01 & 12.31 & 0.01 & 181 & 232 & 177 & B9V\\
15.53 & 0.02 & 14.55 & 0.01 & 13.73 & 0.01 & 12.97 & 0.01 & 12.74 & 0.01 & 12.34 & 0.01 & 249 & 241 & 188 & B9V\\
15.49 & 0.01 & 14.60 & 0.01 & 13.68 & 0.01 & 12.96 & 0.01 & 12.75 & 0.01 & 12.35 & 0.01 & 86 & 63 & 33 & B9V\\
15.58 & 0.01 & 14.60 & 0.01 & 13.78 & 0.01 & 13.01 & 0.01 & 12.81 & 0.01 & 12.41 & 0.01 & 239 & 363 & 302 & B9V\\
16.98 & 0.02 & 16.11 & 0.01 & 15.23 & 0.01 & 13.28 & 0.01 & 12.85 & 0.01 & 11.96 & 0.01 & 147 & 223 & 166 & K0III\\
15.57 & 0.01 & 14.63 & 0.01 & 13.86 & 0.01 & 13.14 & 0.01 & 12.97 & 0.01 & 12.61 & 0.01 & 227 & 440 & 360 & B8V\\
15.87 & 0.01 & 14.91 & 0.01 & 14.03 & 0.01 & 13.22 & 0.01 & 13.00 & 0.01 & 12.60 & 0.01 & 57 & 414 & 340 & B9V\\
15.59 & 0.01 & 14.71 & 0.01 & 13.93 & 0.01 & 13.20 & 0.01 & 13.01 & 0.01 & 12.69 & 0.01 & 284 & 412 & 339 & A0V\\
15.83 & 0.01 & 14.84 & 0.01 & 14.00 & 0.01 & 13.24 & 0.01 & 13.03 & 0.01 & 12.66 & 0.01 & 273 & 315 & 260 & B9V\\
16.09 & 0.01 & 15.10 & 0.01 & 14.28 & 0.01 & 13.30 & 0.01 & 13.07 & 0.01 & 12.61 & 0.02 & 342 & 547 & 415 & B9V\\
16.06 & 0.01 & 15.04 & 0.01 & 14.24 & 0.01 & 13.46 & 0.01 & 13.27 & 0.01 & 12.88 & 0.01 & 162 & 427 & 350 & B9V\\
16.32 & 0.02 & 15.37 & 0.01 & 14.42 & 0.01 & 13.53 & 0.01 & 13.30 & 0.01 & 12.88 & 0.01 & 69 & 395 & 326 & A0V\\
15.95 & 0.01 & 15.11 & 0.01 & 14.25 & 0.01 & 13.52 & 0.01 & 13.32 & 0.01 & 12.96 & 0.01 & 225 & 147 & 97 & B9V\\
16.13 & 0.01 & 15.24 & 0.01 & 14.33 & 0.01 & 13.54 & 0.01 & 13.33 & 0.01 & 12.94 & 0.01 & 54 & 207 & 145 & A0V\\
16.14 & 0.02 & 15.34 & 0.02 & 14.43 & 0.01 & 13.53 & 0.01 & 13.34 & 0.01 & 12.87 & 0.02 & 204 & 184 & 127 & B9V\\
16.18 & 0.01 & 15.31 & 0.01 & 14.41 & 0.01 & 13.62 & 0.01 & 13.42 & 0.01 & 13.06 & 0.01 & 120 & 219 & 148 & A0V\\
16.21 & 0.01 & 15.32 & 0.01 & 14.49 & 0.01 & 13.71 & 0.01 & 13.49 & 0.01 & 13.10 & 0.01 & 178 & 302 & 246 & B9V\\
16.08 & 0.01 & 15.20 & 0.01 & 14.44 & 0.01 & 13.72 & 0.01 & 13.52 & 0.01 & 13.18 & 0.01 & 331 & 265 & 216 & B9V\\
16.29 & 0.02 & 15.46 & 0.02 & 14.60 & 0.01 & 13.84 & 0.01 & 13.56 & 0.01 & 13.23 & 0.01 & 210 & 228 & 171 & A0V\\
16.39 & 0.02 & 15.44 & 0.01 & 14.57 & 0.01 & 13.77 & 0.01 & 13.56 & 0.01 & 13.19 & 0.01 & 216 & 381 & 316 & A0V\\
16.51 & 0.02 & 15.62 & 0.01 & 14.71 & 0.01 & 13.94 & 0.01 & 13.73 & 0.01 & 13.35 & 0.01 & 61 & 340 & 283 & B9V\\
16.48 & 0.01 & 15.58 & 0.01 & 14.77 & 0.01 & 13.95 & 0.01 & 13.75 & 0.01 & 13.36 & 0.01 & 301 & 393 & 325 & B9V\\
16.53 & 0.02 & 15.65 & 0.01 & 14.80 & 0.01 & 13.95 & 0.01 & 13.77 & 0.01 & 13.29 & 0.01 & 127 & 263 & 211 & B9V\\
16.66 & 0.02 & 15.76 & 0.01 & 14.81 & 0.01 & 14.01 & 0.01 & 13.80 & 0.01 & 13.44 & 0.01 & 98 & 149 & 100 & A0V\\
16.78 & 0.02 & 15.82 & 0.01 & 14.88 & 0.01 & 14.09 & 0.01 & 13.88 & 0.01 & 13.52 & 0.01 & 99 & 165 & 110 & A2V\\
16.63 & 0.02 & 15.91 & 0.01 & 14.97 & 0.01 & 14.16 & 0.01 & 13.97 & 0.01 & 13.60 & 0.01 & 162 & 86 & 52 & A0V\\
16.87 & 0.02 & 15.97 & 0.01 & 15.09 & 0.01 & 14.22 & 0.01 & 14.00 & 0.01 & 13.65 & 0.01 & 278 & 298 & 242 & A0V\\
17.06 & 0.02 & 16.12 & 0.02 & 15.19 & 0.01 & 14.27 & 0.02 & 14.01 & 0.02 & 13.58 & 0.02 & 152 & 218 & 158 & A0V\\
16.97 & 0.02 & 16.08 & 0.01 & 15.09 & 0.01 & 14.24 & 0.01 & 14.02 & 0.01 & 13.63 & 0.01 & 20 & 131 & 84 & A0V\\
17.00 & 0.02 & 16.08 & 0.01 & 15.21 & 0.01 & 14.23 & 0.01 & 14.02 & 0.01 & 13.62 & 0.01 & 54 & 523 & 403 & A0V\\
17.39 & 0.04 & 16.58 & 0.02 & 16.02 & 0.01 & 14.39 & 0.01 & 14.04 & 0.01 & 13.69 & 0.01 & 151 & 403 & 330 & G8IV\\
16.88 & 0.01 & 16.05 & 0.01 & 15.11 & 0.01 & 14.27 & 0.01 & 14.07 & 0.01 & 13.71 & 0.01 & 180 & 157 & 106 & A2V\\
16.92 & 0.02 & 16.00 & 0.01 & 15.15 & 0.01 & 14.27 & 0.01 & 14.07 & 0.01 & 13.68 & 0.01 & 322 & 362 & 301 & A0V\\
16.98 & 0.04 & 16.38 & 0.02 & 15.45 & 0.01 & 14.41 & 0.01 & 14.13 & 0.01 & 13.71 & 0.01 & 134 & 215 & 149 & NA\\
17.09 & 0.03 & 16.13 & 0.02 & 15.21 & 0.01 & 14.38 & 0.01 & 14.15 & 0.01 & 13.77 & 0.01 & 162 & 415 & 342 & A0V\\
17.27 & 0.03 & 16.33 & 0.02 & 15.47 & 0.01 & 14.39 & 0.01 & 14.15 & 0.01 & 13.70 & 0.01 & 92 & 328 & 271 & A0V\\
17.35 & 0.03 & 16.49 & 0.02 & 15.82 & 0.01 & 14.50 & 0.01 & 14.16 & 0.01 & 13.50 & 0.01 & 44 & 376 & 311 & K0VI\\
17.20 & 0.02 & 16.24 & 0.02 & 15.35 & 0.01 & 14.40 & 0.01 & 14.18 & 0.01 & 13.75 & 0.01 & 117 & 471 & 379 & A2V\\
17.09 & 0.03 & 16.26 & 0.02 & 15.31 & 0.01 & 14.44 & 0.01 & 14.22 & 0.01 & 13.79 & 0.01 & 69 & 230 & 176 & A5V\\
17.12 & 0.03 & 16.28 & 0.02 & 15.34 & 0.01 & 14.44 & 0.01 & 14.22 & 0.01 & 13.84 & 0.01 & 171 & 292 & 236 & A0V\\
17.41 & 0.02 & 16.57 & 0.02 & 15.98 & 0.01 & 14.60 & 0.01 & 14.27 & 0.01 & 13.60 & 0.01 & 315 & 542 & 412 & K2V\\
18.86 & 0.11 & 18.16 & 0.07 & 17.15 & 0.03 & 14.80 & 0.01 & 14.27 & 0.01 & 13.26 & 0.02 & 214 & 81 & 47 & K0III\\
17.11 & 0.03 & 16.29 & 0.02 & 15.36 & 0.01 & 14.51 & 0.01 & 14.29 & 0.01 & 13.93 & 0.01 & 221 & 170 & 115 & A2V\\
17.31 & 0.03 & 16.37 & 0.03 & 15.44 & 0.01 & 14.55 & 0.01 & 14.31 & 0.01 & 13.96 & 0.01 & 80 & 153 & 102 & NA\\
17.31 & 0.03 & 16.46 & 0.02 & 15.58 & 0.01 & 14.56 & 0.01 & 14.31 & 0.01 & 13.93 & 0.01 & 103 & 316 & 269 & A5V\\
17.45 & 0.03 & 16.44 & 0.02 & 15.55 & 0.01 & 14.50 & 0.01 & 14.39 & 0.01 & 13.98 & 0.01 & 143 & 283 & 227 & A5V\\
17.33 & 0.03 & 16.46 & 0.02 & 15.52 & 0.01 & 14.60 & 0.01 & 14.40 & 0.01 & 14.04 & 0.01 & 13 & 377 & 310 & A5V\\
17.23 & 0.03 & 16.46 & 0.02 & 15.50 & 0.01 & 14.65 & 0.01 & 14.43 & 0.01 & 14.07 & 0.01 & 248 & 183 & 128 & A2V\\
17.32 & 0.03 & 16.45 & 0.04 & 15.74 & 0.01 & 14.74 & 0.01 & 14.44 & 0.01 & 13.99 & 0.01 & 227 & 232 & 179 & A5V\\
17.36 & 0.03 & 16.60 & 0.02 & 15.62 & 0.01 & 14.70 & 0.01 & 14.47 & 0.01 & 14.05 & 0.01 & 131 & 105 & 62 & A5V\\
17.52 & 0.03 & 16.75 & 0.02 & 15.73 & 0.01 & 14.78 & 0.01 & 14.55 & 0.01 & 14.15 & 0.01 & 39 & 80 & 46 & A2V\\
17.63 & 0.03 & 16.70 & 0.02 & 15.84 & 0.01 & 14.83 & 0.01 & 14.58 & 0.01 & 14.21 & 0.01 & 116 & 489 & 387 & NA\\
17.58 & 0.03 & 16.70 & 0.03 & 15.82 & 0.01 & 14.83 & 0.01 & 14.60 & 0.01 & 14.16 & 0.01 & 265 & 472 & 380 & A0V\\
17.67 & 0.04 & 16.81 & 0.02 & 15.86 & 0.01 & 14.77 & 0.01 & 14.60 & 0.01 & 14.18 & 0.01 & 163 & 220 & 162 & A5V\\
17.64 & 0.04 & 16.73 & 0.03 & 15.82 & 0.01 & 14.82 & 0.01 & 14.62 & 0.01 & 14.20 & 0.01 & 92 & 338 & 281 & A5V\\
17.58 & 0.04 & 16.71 & 0.03 & 15.84 & 0.01 & 14.89 & 0.01 & 14.63 & 0.01 & 14.24 & 0.01 & 218 & 185 & 130 & A2V\\
17.59 & 0.03 & 16.77 & 0.03 & 15.85 & 0.01 & 14.86 & 0.01 & 14.63 & 0.01 & 14.22 & 0.01 & 219 & 216 & 157 & A5V\\
17.50 & 0.04 & 16.72 & 0.02 & 15.84 & 0.01 & 14.89 & 0.01 & 15.65 & 0.01 & 14.24 & 0.01 & 294 & 71 & 39 & A5V\\
17.66 & 0.03 & 16.74 & 0.02 & 15.87 & 0.01 & 14.87 & 0.01 & 14.65 & 0.01 & 14.22 & 0.01 & 208 & 269 & 217 & A5V\\
17.78 & 0.05 & 17.14 & 0.03 & 16.48 & 0.02 & 15.02 & 0.01 & 14.65 & 0.01 & 13.97 & 0.01 & 52 & 44 & 15 & K4V\\
17.66 & 0.04 & 16.96 & 0.05 & 16.38 & 0.05 & 15.08 & 0.06 & 14.68 & 0.06 & 13.94 & 0.03 & 317 & 221 & 164 & G2V\\
17.74 & 0.04 & 16.76 & 0.03 & 15.95 & 0.01 & 14.99 & 0.01 & 14.74 & 0.01 & 14.33 & 0.01 & 279 & 250 & 193 & A5V\\
17.86 & 0.04 & 16.88 & 0.02 & 15.95 & 0.01 & 14.99 & 0.01 & 14.75 & 0.01 & 14.36 & 0.01 & 124 & 384 & 318 & A5V\\
17.65 & 0.03 & 16.85 & 0.02 & 16.00 & 0.01 & 15.01 & 0.01 & 14.76 & 0.01 & 14.34 & 0.01 & 285 & 178 & 123 & A5V\\
17.87 & 0.05 & 17.04 & 0.04 & 16.04 & 0.02 & 15.06 & 0.01 & 14.76 & 0.03 & 14.42 & 0.02 & 171 & 310 & 254 & NA\\
17.79 & 0.04 & 16.95 & 0.03 & 16.01 & 0.01 & 15.01 & 0.01 & 14.78 & 0.01 & 14.36 & 0.01 & 157 & 375 & 308 & A5V\\
17.68 & 0.03 & 16.97 & 0.02 & 16.05 & 0.01 & 15.06 & 0.01 & 14.82 & 0.01 & 14.39 & 0.01 & 252 & 107 & 65 & A5V\\
17.74 & 0.03 & 17.01 & 0.03 & 16.32 & 0.01 & 15.11 & 0.01 & 14.82 & 0.01 & 14.26 & 0.01 & 333 & 105 & 64 & F0V\\
17.88 & 0.04 & 16.99 & 0.03 & 16.20 & 0.01 & 15.04 & 0.02 & 14.82 & 0.02 & 14.40 & 0.02 & 210 & 245 & 191 & A5V\\
17.87 & 0.04 & 17.05 & 0.03 & 16.12 & 0.01 & 15.11 & 0.01 & 14.85 & 0.01 & 14.42 & 0.01 & 208 & 286 & 234 & A5V\\
17.98 & 0.05 & 17.12 & 0.04 & 16.17 & 0.01 & 15.17 & 0.01 & 14.88 & 0.01 & 14.46 & 0.01 & 44 & 202 & 143 & A5V\\
17.86 & 0.04 & 16.99 & 0.02 & 16.14 & 0.01 & 15.16 & 0.01 & 14.91 & 0.01 & 14.51 & 0.01 & 268 & 234 & 182 & A5V\\
17.83 & 0.04 & 17.01 & 0.04 & 16.33 & 0.03 & 15.22 & 0.03 & 14.95 & 0.02 & 14.55 & 0.01 & 294 & 282 & 226 & A5V\\
18.04 & 0.04 & 17.04 & 0.03 & 16.25 & 0.01 & 15.21 & 0.01 & 14.96 & 0.01 & 14.55 & 0.01 & 236 & 534 & 407 & K1IV\\
18.17 & 0.06 & 17.17 & 0.03 & 16.30 & 0.01 & 15.28 & 0.01 & 15.02 & 0.01 & 14.63 & 0.01 & 123 & 357 & 298 & A5V\\
17.32 & 0.06 & 16.44 & 0.04 & 16.36 & 0.01 & 15.26 & 0.01 & 15.05 & 0.01 & 14.59 & 0.01 & 231 & 235 & 183 & NA\\
18.27 & 0.07 & 17.27 & 0.03 & 16.40 & 0.02 & 15.32 & 0.01 & 15.05 & 0.01 & 14.60 & 0.01 & 108 & 256 & 200 & A5V\\
18.03 & 0.05 & 17.22 & 0.04 & 16.33 & 0.01 & 15.34 & 0.01 & 15.06 & 0.01 & 14.63 & 0.01 & 237 & 259 & 207 & A5V\\
18.28 & 0.07 & 17.40 & 0.03 & 16.53 & 0.01 & 15.36 & 0.01 & 15.07 & 0.01 & 14.60 & 0.01 & 194 & 213 & 153 & A5V\\
18.24 & 0.05 & 17.29 & 0.03 & 16.44 & 0.01 & 15.39 & 0.01 & 15.12 & 0.01 & 14.71 & 0.01 & 283 & 465 & 376 & A5V\\
18.25 & 0.05 & 17.49 & 0.05 & 16.71 & 0.03 & 15.43 & 0.03 & 15.17 & 0.03 & 14.47 & 0.04 & 301 & 75 & 43 & F5V\\
18.41 & 0.06 & 17.45 & 0.03 & 16.64 & 0.02 & 15.49 & 0.01 & 15.19 & 0.01 & 14.78 & 0.01 & 203 & 321 & 267 & NA\\
18.58 & 0.08 & 17.46 & 0.03 & 16.64 & 0.02 & 15.54 & 0.01 & 15.28 & 0.01 & 14.89 & 0.02 & 145 & 233 & 181 & NA\\
18.33 & 0.05 & 17.53 & 0.04 & 16.63 & 0.02 & 15.57 & 0.01 & 15.29 & 0.01 & 14.82 & 0.01 & 97 & 176 & 119 & A5V\\
18.41 & 0.07 & 17.44 & 0.04 & 16.71 & 0.02 & 15.62 & 0.01 & 15.32 & 0.01 & 14.83 & 0.01 & 339 & 342 & 288 & A7V\\
18.43 & 0.06 & 17.63 & 0.05 & 16.73 & 0.02 & 15.59 & 0.01 & 15.33 & 0.01 & 14.86 & 0.01 & 95 & 208 & 147 & A5V\\
18.38 & 0.07 & 17.52 & 0.04 & 16.81 & 0.02 & 15.63 & 0.01 & 15.40 & 0.01 & 14.91 & 0.01 & 185 & 541 & 411 & A7V\\
18.43 & 0.07 & 17.56 & 0.04 & 16.79 & 0.01 & 15.68 & 0.01 & 15.40 & 0.01 & 14.93 & 0.01 & 245 & 342 & 287 & A7V\\
18.37 & 0.06 & 17.57 & 0.04 & 17.07 & 0.02 & 15.79 & 0.01 & 15.46 & 0.01 & 14.86 & 0.01 & 316 & 298 & 243 & F0V\\
18.75 & 0.08 & 17.69 & 0.04 & 17.08 & 0.02 & 15.86 & 0.02 & 15.53 & 0.02 & 15.08 & 0.02 & 301 & 413 & 341 & NA\\
18.39 & 0.05 & 17.81 & 0.05 & 17.23 & 0.03 & 15.96 & 0.02 & 15.59 & 0.01 & 14.98 & 0.01 & 354 & 62 & 32 & F2V\\
18.68 & 0.08 & 17.65 & 0.04 & 17.07 & 0.02 & 15.94 & 0.01 & 15.61 & 0.01 & 15.09 & 0.01 & 331 & 251 & 194 & F0V\\
19.32 & 0.21 & 18.06 & 0.07 & 17.83 & 0.03 & 16.03 & 0.01 & 15.69 & 0.02 & 15.12 & 0.01 & 162 & 241 & 187 & NA\\
18.60 & 0.08 & 17.89 & 0.05 & 17.18 & 0.02 & 16.01 & 0.01 & 15.71 & 0.01 & 15.16 & 0.01 & 325 & 335 & 279 & F0V\\
18.94 & 0.11 & 18.08 & 0.07 & 17.40 & 0.03 & 16.05 & 0.03 & 15.71 & 0.04 & 15.39 & 0.01 & 270 & 134 & 91 & NA\\
18.81 & 0.09 & 17.86 & 0.06 & 17.25 & 0.03 & 16.04 & 0.01 & 15.72 & 0.01 & 15.15 & 0.01 & 296 & 102 & 59 & A7V\\
18.64 & 0.08 & 18.03 & 0.06 & 17.49 & 0.03 & 16.08 & 0.02 & 15.74 & 0.02 & 15.18 & 0.01 & 103 & 58 & 26 & A7V\\
18.80 & 0.11 & 18.19 & 0.06 & 17.49 & 0.03 & 16.12 & 0.02 & 15.78 & 0.01 & 15.13 & 0.01 & 19 & 62 & 31 & F0V\\
19.01 & 0.10 & 18.24 & 0.07 & 17.40 & 0.03 & 16.18 & 0.01 & 15.88 & 0.01 & 15.29 & 0.01 & 196 & 347 & 293 & A7V\\
19.25 & 0.16 & 18.43 & 0.08 & 17.65 & 0.03 & 16.34 & 0.02 & 15.96 & 0.02 & 15.37 & 0.01 & 43 & 333 & 277 & NA\\
19.10 & 0.17 & 18.25 & 0.06 & 17.58 & 0.03 & 16.29 & 0.02 & 16.01 & 0.01 & 15.43 & 0.02 & 115 & 548 & 416 & A7V\\
18.89 & 0.09 & 18.24 & 0.07 & 17.71 & 0.03 & 16.46 & 0.01 & 16.09 & 0.01 & 15.51 & 0.01 & 362 & 154 & 103 & F2V\\
18.77 & 0.10 & 18.50 & 0.09 & 17.70 & 0.03 & 16.45 & 0.01 & 16.14 & 0.01 & 15.53 & 0.01 & 265 & 483 & 385 & F0V\\
19.50 & 0.17 & 18.66 & 0.10 & 17.90 & 0.03 & 16.49 & 0.03 & 16.24 & 0.03 & 15.68 & 0.02 & 299 & 314 & 261 & NA\\
18.79 & 0.08 & 17.96 & 0.07 & 17.66 & 0.05 & 16.21 & 0.04 & 16.67 & 0.41 & 16.04 & 0.04 & 332 & 6 & 2 & NA\\
\end{longtable}

\clearpage
\onecolumn

\clearpage
\onecolumn

\begin{longtable}{cccccccccccc}
\caption{\color{black}{\bf Gaia DR2 data for the stars in direction of the Melotte 105:} ID is the identification assigned to a star by this study, $\alpha$ is right ascension, $\delta$ declination, `Gmag' the {\em Gaia} magnitude, `Plx' parallax, $\epsilon(\rm Plx)$ the error in the parallax, $\alpha_{pm}$ right ascension proper motion, $\epsilon (\alpha_{pm})$ error in the right ascension proper motion, $\delta_{pm} $ declination proper motion, $\epsilon (\delta_{pm})$ error in the declination proper motion, and `Prob' the probability of cluster membership derived by this study. BP stands for the {\em Gaia} Blue Photometer, which operates in the wavelength range 330–680 nm. RP stands for the {\em Gaia} Red Photometer, which covers the wavelength range 640–1050 nm, hence `BP-RP' is the colour difference between these two filters.
\label{tab:gaia}}\\
 \hline
 ID & $\alpha$ & $\delta $ & $G$ & $BP-RP$ & Plx &$\bf \epsilon(\rm Plx)$ &  $\alpha_{pm}$ & $\epsilon (\alpha_{pm})$ & $\delta_{pm}$ & $\epsilon (\delta_{pm}) $ & Prob \\
\hline
\endfirsthead
\multicolumn{12}{c}%
{\tablename\ \thetable\ -- \textit{Continued from previous page}} \\
 \hline
 ID & $\alpha$  & $\delta $ & Gmag &
 BP$-$RP & Plx &$\bf \epsilon(\rm Plx)$ &
 $\alpha_{pm}$ & $\epsilon ( \alpha_{pm} ) $ & $\delta_{pm} $ &
$\epsilon ( \delta_{pm} ) $ & Prob \\
\hline
 \endhead
  184 &   11:19:36.94 & -63:29:11.27 & 11.316 &  1.679 &   0.3843 &  0.0282 & -6.750 &  0.050 &  2.150 &  0.045 &   0.97 \\
  201 &   11:19:41.54 & -63:28:59.15 & 11.812 &  0.658 &   0.1348 &  0.0509 & -6.120 &  0.085 &  1.897 &  0.073 &   0.00 \\
  258 &   11:19:45.29 & -63:28:24.61 & 11.758 &  0.710 &   0.4109 &  0.0262 & -6.755 &  0.042 &  1.978 &  0.045 &   0.99 \\
  139 &   11:19:53.28 & -63:29:36.79 & 11.549 &  1.649 &   0.4444 &  0.0331 & -6.716 &  0.051 &  2.186 &  0.048 &   0.97 \\
  358 &   11:19:27.95 & -63:27:07.18 & 11.802 &  1.593 &   0.4151 &  0.0263 & -6.885 &  0.042 &  2.192 &  0.043 &   0.98 \\
  116 &   11:19:45.07 & -63:29:53.10 & 12.282 &  0.622 &   0.4037 &  0.0315 & -6.903 &  0.055 &  2.088 &  0.066 &   0.98 \\
  190 &   11:19:38.29 & -63:29:07.40 & 12.383 &  0.654 &   0.4038 &  0.0280 & -6.717 &  0.046 &  2.136 &  0.043 &   0.95 \\
  117 &   11:19:44.73 & -63:29:52.64 & 12.282 &  0.622 &   0.4037 &  0.0315 & -6.903 &  0.055 &  2.088 &  0.066 &   0.95 \\
  296 &   11:19:30.72 & -63:27:58.82 & 12.425 &  0.594 &   0.4280 &  0.0281 & -6.890 &  0.045 &  2.095 &  0.047 &   0.95 \\
  266 &   11:19:48.25 & -63:28:20.62 & 12.506 &  0.660 &   0.3818 &  0.0290 & -6.722 &  0.047 &  2.597 &  0.048 &   0.47 \\
  259 &   11:19:38.59 & -63:28:23.82 & 12.594 &  0.701 &   0.3613 &  0.0284 & -6.737 &  0.045 &  2.056 &  0.048 &   0.97 \\
  177 &   11:19:39.07 & -63:29:14.80 & 12.651 &  0.669 &   0.3764 &  0.0298 & -6.735 &  0.052 &  2.308 &  0.054 &   0.84 \\
  188 &   11:19:32.69 & -63:29:08.76 & 12.686 &  0.645 &   0.4118 &  0.0289 & -6.786 &  0.049 &  2.190 &  0.045 &   0.97 \\
   33 &   11:19:47.76 & -63:30:58.31 & 12.751 &  0.614 &   0.4192 &  0.0280 & -6.880 &  0.040 &  2.080 &  0.040 &   0.99 \\
  302 &   11:19:33.81 & -63:27:53.67 & 12.760 &  0.650 &   0.4897 &  0.0307 & -6.938 &  0.048 &  2.199 &  0.047 &   0.97 \\
  166 &   11:19:42.10 & -63:29:21.51 & 12.530 &  1.549 &    ---   &   ---   &   --   &  ---   &   ---  &   ---  &   0.00 \\
  360 &   11:19:35.01 & -63:27:05.67 & 12.942 &  0.544 &   0.4318 &  0.0228 & -6.874 &  0.034 &  2.261 &  0.037 &   1.00 \\
  340 &   11:19:50.75 & -63:27:23.00 & 12.941 &  0.653 &   0.4136 &  0.0206 & -6.767 &  0.033 &  2.083 &  0.031 &   1.00 \\
  339 &   11:19:29.66 & -63:27:22.40 & 12.997 &  0.534 &   0.4301 &  0.0208 & -6.803 &  0.033 &  2.068 &  0.032 &   1.00 \\
  260 &   11:19:30.56 & -63:28:22.98 & 13.015 &  0.633 &   0.3503 &  0.0204 & -6.748 &  0.033 &  2.036 &  0.033 &   0.95 \\
  415 &   11:19:24.68 & -63:25:57.27 & 13.010 &  0.777 &   0.5366 &  0.0233 & -9.468 &  0.034 &  3.383 &  0.036 &   0.32 \\
  350 &   11:19:41.06 & -63:27:13.90 & 13.226 &  0.615 &   0.4074 &  0.0179 & -6.749 &  0.030 &  1.842 &  0.028 &   0.81 \\
  326 &   11:19:49.53 & -63:27:35.06 & 13.249 &  0.722 &   0.6299 &  0.0166 & -4.543 &  0.027 &  1.764 &  0.026 &   0.22 \\
   97 &   11:19:34.79 & -63:30:07.11 & 13.316 &  0.557 &   0.3940 &  0.0173 & -6.890 &  0.020 &  1.940 &  0.020 &   0.87 \\
  145 &   11:19:50.66 & -63:29:30.99 & 13.283 &  0.639 &   0.4004 &  0.0187 & -7.267 &  0.029 &  2.179 &  0.029 &   0.53 \\
  127 &   11:19:36.93 & -63:29:44.20 & 13.464 &  0.691 &    ---   &   ---   &   --   &  ---   &   ---  &   ---  &   0.00 \\
  148 &   11:19:44.63 & -63:29:29.66 & 13.384 &  0.613 &   0.4041 &  0.0183 & -6.618 &  0.032 &  2.003 &  0.031 &   0.88 \\
  246 &   11:19:39.48 & -63:28:31.20 & 13.460 &  0.631 &   0.3894 &  0.0172 & -6.710 &  0.028 &  1.837 &  0.027 &   0.76 \\
  216 &   11:19:25.28 & -63:28:51.00 & 13.504 &  0.539 &   0.3833 &  0.0169 & -6.723 &  0.026 &  2.197 &  0.026 &   0.97 \\
  171 &   11:19:36.37 & -63:29:17.10 & 13.552 &  0.646 &   0.4219 &  0.0208 & -6.822 &  0.034 &  2.078 &  0.032 &   0.97 \\
  316 &   11:19:36.01 & -63:27:42.10 & 13.512 &  0.636 &   0.4043 &  0.0186 & -6.972 &  0.029 &  2.047 &  0.030 &   0.98 \\
  283 &   11:19:50.24 & -63:28:08.40 & 13.670 &  0.620 &   0.3774 &  0.0201 & -6.565 &  0.031 &  2.248 &  0.029 &   0.85 \\
  325 &   11:19:28.08 & -63:27:34.21 & 13.704 &  0.632 &   0.3851 &  0.0223 & -6.753 &  0.032 &  1.980 &  0.031 &   0.96 \\
  211 &   11:19:44.11 & -63:28:55.30 & 13.650 &  0.679 &   0.4379 &  0.0188 & -6.759 &  0.030 &  2.055 &  0.035 &   0.95 \\
  100 &   11:19:46.52 & -63:30:06.87 & 13.779 &  0.610 &   0.4375 &  0.0198 & -6.807 &  0.032 &  2.032 &  0.030 &   1.00 \\
  110 &   11:19:46.55 & -63:29:56.60 & 13.855 &  0.620 &   0.4258 &  0.0229 & -6.761 &  0.038 &  2.166 &  0.039 &   0.98 \\
   52 &   11:19:40.56 & -63:30:45.07 & 13.948 &  0.631 &   0.4702 &  0.0222 & -6.540 &  0.030 &  2.200 &  0.030 &   0.97 \\
  242 &   11:19:30.19 & -63:28:33.40 & 13.981 &  0.679 &   0.5363 &  0.0216 & -6.966 &  0.032 &  2.307 &  0.032 &   0.90 \\
  158 &   11:19:41.74 & -63:29:23.50 & 14.075 &  0.704 &   0.3468 &  0.0252 & -6.796 &  0.039 &  2.319 &  0.041 &   0.99 \\
   84 &   11:19:53.70 & -63:30:18.69 & 13.991 &  0.675 &   0.3954 &  0.0199 & -6.740 &  0.030 &  2.180 &  0.030 &   0.98 \\
  403 &   11:19:51.07 & -63:26:14.10 & 13.979 &  0.710 &   0.5436 &  0.0209 & -4.447 &  0.032 &  0.167 &  0.030 &   0.18 \\
  330 &   11:19:41.91 & -63:27:29.44 & 13.818 &  1.313 &   1.0824 &  0.0244 &  2.941 &  0.040 & -3.973 &  0.039 &   0.00 \\
  106 &   11:19:38.99 & -63:30:01.33 & 14.022 &  0.629 &   0.3826 &  0.0205 & -6.794 &  0.036 &  1.898 &  0.028 &   0.91 \\
  301 &   11:19:26.14 & -63:27:53.50 & 14.049 &  0.655 &   0.4006 &  0.0218 & -6.729 &  0.033 &  2.037 &  0.038 &   0.99 \\
  149 &   11:19:43.37 & -63:29:25.60 & 14.084 &  0.850 &   0.3968 &  0.0234 & -6.715 &  0.035 &  1.934 &  0.038 &   0.99 \\
  271 &   11:19:47.35 & -63:28:16.20 & 14.198 &  0.823 &   0.3420 &  0.0213 & -6.683 &  0.033 &  2.080 &  0.033 &   0.77 \\
  342 &   11:19:40.95 & -63:27:21.77 & 14.140 &  0.652 &   0.4086 &  0.0219 & -6.783 &  0.037 &  2.028 &  0.034 &   0.96 \\
  311 &   11:19:51.81 & -63:27:46.61 & 13.963 &  1.160 &   0.5115 &  0.0202 & -14.180&  0.034 &  3.393 &  0.032 &   0.23 \\
  379 &   11:19:45.32 & -63:26:47.23 & 14.092 &  0.743 &   0.3908 &  0.0193 & -6.371 &  0.030 &  1.842 &  0.031 &   0.32 \\
  176 &   11:19:49.39 & -63:29:16.40 & 14.242 &  0.683 &   0.3410 &  0.0204 & -6.747 &  0.035 &  2.122 &  0.036 &   0.93 \\
  236 &   11:19:39.92 & -63:28:37.50 & 14.201 &  0.696 &   0.3586 &  0.0262 & -6.985 &  0.043 &  2.172 &  0.035 &   0.96 \\
   47 &   11:19:35.88 & -63:30:47.47 & 13.854 &  1.780 &   0.3622 &  0.0187 & -9.230 &  0.020 &  2.860 &  0.020 &   0.41 \\
  412 &   11:19:27.10 & -63:26:00.22 & 14.093 &  1.152 &   0.7124 &  0.0365 & -11.390&  0.057 &  3.411 &  0.055 &   0.19 \\
  115 &   11:19:35.18 & -63:29:52.99 & 14.267 &  0.646 &   0.3843 &  0.0216 & -6.668 &  0.035 &  2.188 &  0.030 &   0.99 \\
  102 &   11:19:48.30 & -63:30:03.90 & 14.343 &  0.663 &   0.3305 &  0.0254 & -6.897 &  0.042 &  2.111 &  0.039 &   0.91 \\
  269 &   11:19:46.29 & -63:28:17.28 & 14.286 &  0.749 &   0.4296 &  0.0204 & -6.854 &  0.032 &  2.098 &  0.034 &   0.96 \\
  227 &   11:19:42.59 & -63:28:43.20 & 14.316 &  0.744 &   0.4349 &  0.0205 & -6.720 &  0.034 &  2.146 &  0.032 &   0.95 \\
  310 &   11:19:54.65 & -63:27:46.66 & 14.345 &  0.681 &   0.4030 &  0.0213 & -6.722 &  0.035 &  2.104 &  0.038 &   1.00 \\
  128 &   11:19:32.69 & -63:29:44.67 & 14.393 &  0.650 &   0.3917 &  0.0234 & -6.708 &  0.035 &  2.193 &  0.033 &   0.96 \\
  179 &   11:19:34.87 & -63:29:14.20 & 14.348 &  0.863 &   0.3861 &  0.0227 & -6.692 &  0.036 &  2.219 &  0.032 &   0.96 \\
   62 &   11:19:43.47 & -63:30:34.03 & 14.412 &  0.728 &   0.3836 &  0.0215 & -6.760 &  0.030 &  2.200 &  0.030 &   1.00 \\
   46 &   11:19:51.54 & -63:30:43.84 &  ---   &   ---  &    ---   &   ---   &   ---  &   ---  &   ---  &   ---  &   0.00 \\
  387 &   11:19:45.32 & -63:26:34.59 & 14.545 &  0.760 &   0.4600 &  0.0217 & -7.048 &  0.034 &  1.899 &  0.036 &   0.71 \\
  162 &   11:19:40.74 & -63:29:21.90 & 14.512 &  0.806 &   0.3608 &  0.0320 & -6.986 &  0.050 &  1.860 &  0.043 &   0.94 \\
  380 &   11:19:31.76 & -63:26:45.41 & 14.519 &  0.766 &   0.3754 &  0.0227 & -6.719 &  0.037 &  2.281 &  0.036 &   0.96 \\
  281 &   11:19:47.27 & -63:28:09.65 & 14.555 &  0.801 &   0.4032 &  0.0225 & -6.673 &  0.036 &  2.179 &  0.034 &   0.98 \\
  130 &   11:19:35.53 & -63:29:43.71 & 14.571 &  0.757 &   0.4056 &  0.0239 & -6.662 &  0.037 &  2.186 &  0.033 &   0.95 \\
  157 &   11:19:35.48 & -63:29:24.76 & 14.570 &  0.763 &   0.3872 &  0.0251 & -6.801 &  0.038 &  2.390 &  0.035 &   0.90 \\
   15 &   11:19:50.82 & -63:31:11.54 & 14.505 &  1.185 &   1.1030 &  0.0206 & -16.580&  0.033 & -0.338 &  0.032 &   0.00 \\
  217 &   11:19:36.62 & -63:28:51.40 & 14.577 &  0.753 &   0.4033 &  0.0244 & -6.774 &  0.037 &  2.040 &  0.036 &   0.96 \\
  164 &   11:19:26.53 & -63:29:20.12 & 14.955 &  1.232 &   0.4912 &  0.0265 & -7.624 &  0.042 & -0.213 &  0.043 &   0.44 \\
  193 &   11:19:30.04 & -63:29:02.50 & 14.686 &  0.722 &   0.3796 &  0.0250 & -6.569 &  0.042 &  2.016 &  0.056 &   0.79 \\
  318 &   11:19:44.37 & -63:27:40.98 & 14.662 &  0.740 &   0.3918 &  0.0229 & -6.854 &  0.037 &  2.121 &  0.035 &   0.99 \\
  123 &   11:19:29.29 & -63:29:47.41 & 14.689 &  0.769 &   0.3710 &  0.0236 & -6.842 &  0.037 &  2.239 &  0.035 &   0.99 \\
  254 &   11:19:40.01 & -63:28:26.86 & 14.807 &  0.755 &   0.4118 &  0.0259 & -6.793 &  0.039 &  2.201 &  0.045 &   0.95 \\
  308 &   11:19:41.30 & -63:27:46.80 & 14.679 &  0.757 &   0.4427 &  0.0244 & -6.795 &  0.040 &  2.268 &  0.038 &   0.97 \\
   64 &   11:19:24.76 & -63:30:32.50 & 14.719 &  0.953 &   0.4388 &  0.0240 & -6.650 &  0.030 &  2.190 &  0.030 &   0.96 \\
   65 &   11:19:32.29 & -63:30:31.39 & 14.748 &  0.775 &   0.4324 &  0.0242 & -6.730 &  0.030 &  1.960 &  0.030 &   0.96 \\
  191 &   11:19:36.29 & -63:29:06.31 & 14.795 &  0.801 &   0.4068 &  0.0254 & -6.839 &  0.043 &  2.193 &  0.043 &   1.00 \\
  234 &   11:19:36.54 & -63:28:40.33 & 14.752 &  0.796 &   0.3827 &  0.0273 & -6.787 &  0.041 &  2.293 &  0.039 &   1.00 \\
  143 &   11:19:51.55 & -63:29:34.11 & 14.824 &  0.814 &   0.3991 &  0.0281 & -6.729 &  0.043 &  2.258 &  0.047 &   0.95 \\
  182 &   11:19:30.91 & -63:29:13.19 & 14.822 &  0.747 &   0.4010 &  0.0282 & -6.770 &  0.041 &  2.245 &  0.040 &   0.98 \\
  226 &   11:19:28.65 & -63:28:43.20 & 14.983 &  0.810 &   0.3543 &  0.0269 & -6.776 &  0.043 &  1.983 &  0.051 &   0.94 \\
  407 &   11:19:34.28 & -63:26:06.88 & 14.886 &  0.796 &   0.4093 &  0.0262 & -6.645 &  0.039 &  1.831 &  0.040 &   0.59 \\
  298 &   11:19:44.42 & -63:27:58.21 & 14.911 &  0.797 &   0.4398 &  0.0303 & -6.769 &  0.056 &  2.141 &  0.043 &   0.98 \\
  183 &   11:19:34.50 & -63:29:12.20 & 14.946 &  0.830 &   0.4317 &  0.0267 & -6.817 &  0.041 &  2.130 &  0.039 &   0.99 \\
  200 &   11:19:45.81 & -63:29:00.00 & 14.944 &  0.859 &   0.3686 &  0.0268 & -6.590 &  0.044 &  2.270 &  0.041 &   0.87 \\
  207 &   11:19:33.80 & -63:28:57.55 & 14.940 &  0.779 &   0.4323 &  0.0272 & -6.784 &  0.047 &  2.169 &  0.051 &   0.98 \\
  153 &   11:19:37.85 & -63:29:26.10 & 14.928 &  0.920 &   0.4483 &  0.0254 & -6.747 &  0.042 &  2.192 &  0.038 &   0.96 \\
  376 &   11:19:30.06 & -63:26:49.61 & 15.028 &  0.809 &   0.4195 &  0.0272 & -6.587 &  0.044 &  2.104 &  0.042 &   1.00 \\
   43 &   11:19:27.70 & -63:30:51.53 & 15.250 &  1.072 &   0.2375 &  0.2339 & -10.390&  0.363 &  4.567 &  0.322 &   0.60 \\
  267 &   11:19:37.05 & -63:28:19.58 & 15.139 &  0.884 &   0.4390 &  0.0298 & -6.735 &  0.049 &  2.168 &  0.044 &   1.00 \\
  181 &   11:19:42.29 & -63:29:14.22 & 15.059 &  0.894 &   0.4001 &  0.0269 & -7.440 &  0.044 &  2.623 &  0.040 &   0.51 \\
  119 &   11:19:46.64 & -63:29:49.91 & 15.191 &  0.869 &   0.3579 &  0.0296 & -6.967 &  0.051 &  2.150 &  0.050 &   0.90 \\
  288 &   11:19:24.47 & -63:28:05.68 & 15.204 &  0.871 &   0.3937 &  0.0310 & -6.555 &  0.048 &  2.115 &  0.046 &   0.92 \\
  147 &   11:19:46.87 & -63:29:30.38 & 15.207 &  0.916 &   0.3967 &  0.0302 & -6.948 &  0.049 &  1.956 &  0.059 &   0.95 \\
  287 &   11:19:33.19 & -63:28:06.13 & 15.260 &  0.875 &   0.4013 &  0.0386 & -6.819 &  0.052 &  2.163 &  0.050 &   0.96 \\
  411 &   11:19:39.13 & -63:26:02.68 & 15.273 &  0.888 &   0.4496 &  0.0296 & -6.871 &  0.045 &  2.257 &  0.047 &   0.95 \\
  243 &   11:19:26.68 & -63:28:33.10 & 15.281 &  1.098 &   0.5288 &  0.0287 & -7.622 &  0.046 &  2.032 &  0.050 &   0.07 \\
  341 &   11:19:28.22 & -63:27:21.60 & 15.489 &  0.940 &   0.4169 &  0.0360 & -6.634 &  0.054 &  2.098 &  0.051 &   0.97 \\
   32 &   11:19:22.92 & -63:30:57.63 & 15.445 &  1.099 &   0.4736 &  0.0362 & -6.550 &  0.050 &  2.480 &  0.040 &   0.86 \\
  194 &   11:19:25.12 & -63:29:02.20 & 15.511 &  0.939 &   0.4688 &  0.0340 & -6.564 &  0.053 &  2.159 &  0.050 &   1.00 \\
   39 &   11:19:28.50 & -63:30:53.46 & 15.250 &  1.072 &   0.2375 &  0.2339 & -10.390&  0.363 &  4.567 &  0.322 &   0.29 \\
  187 &   11:19:40.67 & -63:29:09.29 & 15.433 &  1.101 &   0.3932 &  0.0347 & -6.471 &  0.062 &  2.224 &  0.048 &   0.96 \\
   91 &   11:19:30.66 & -63:30:13.73 & 15.815 &  1.047 &   0.4044 &  0.0381 & -6.710 &  0.060 &  2.360 &  0.050 &   0.93 \\
  279 &   11:19:25.84 & -63:28:09.80 & 15.556 &  0.958 &   0.4100 &  0.0368 & -6.876 &  0.056 &  1.981 &  0.055 &   0.85 \\
   59 &   11:19:28.28 & -63:30:34.20 & 15.569 &  1.008 &   0.4677 &  0.0354 & -6.690 &  0.050 &  1.980 &  0.040 &   0.88 \\
   26 &   11:19:46.14 & -63:31:02.52 & 15.607 &  1.000 &   0.5296 &  0.0329 & -6.770 &  0.050 &  2.160 &  0.040 &   0.95 \\
   31 &   11:19:53.79 & -63:31:00.09 & 15.580 &  1.142 &   0.4031 &  0.0325 & -6.530 &  0.050 &  2.220 &  0.040 &   0.97 \\
  293 &   11:19:37.73 & -63:28:03.70 & 15.687 &  1.039 &   0.3921 &  0.0417 & -6.729 &  0.060 &  2.066 &  0.060 &   0.98 \\
  277 &   11:19:51.77 & -63:28:13.37 & 15.770 &  1.096 &   0.4285 &  0.0385 & -6.621 &  0.060 &  2.208 &  0.058 &   0.96 \\
  416 &   11:19:45.50 & -63:25:58.46 & 15.836 &  1.056 &   0.4587 &  0.0400 & -6.737 &  0.061 &  2.213 &  0.061 &   0.96 \\
  103 &   11:19:22.30 & -63:30:00.43 & 15.941 &  1.060 &   0.5180 &  0.0423 & -6.423 &  0.068 &  2.154 &  0.061 &   0.98 \\
  385 &   11:19:31.58 & -63:26:37.58 & 15.907 &  1.090 &   0.4668 &  0.0397 & -6.683 &  0.064 &  2.197 &  0.061 &   0.97 \\
  261 &   11:19:28.22 & -63:28:23.20 & 16.079 &  1.090 &   0.3947 &  0.0436 & -6.744 &  0.070 &  2.156 &  0.065 &   0.96 \\
   2  &   11:19:24.62 & -63:31:22.93 & 16.851&   1.462 &   0.4119 &  0.0689 &-10.260 &  0.100 &  4.180 &  0.090 &   0.33 \\
\end{longtable}
\clearpage
\onecolumn

 \end{document}